\documentclass[useAMS,usenatbib,fleqn]{mn2e}

\def\physrep{PHYS REP}

\def\mnras{MNRAS}
\def\apj{Ap.J}
\def\aj{AJ}

\def\apjs{Ap.JS}

\def\aap{A \& A}
\def\araa{ARA \& A} 
\def\prd{Phys. Rev. D}

\usepackage[pdftex]{graphicx}
\usepackage{amsmath,amssymb}
\usepackage{times}
\usepackage{color}
\usepackage[T1]{fontenc}
\usepackage{currvita}

\newcommand{\bd}{\begin{displaymath}}
\newcommand{\ed}{\end{displaymath}}
\newcommand{\be}{\begin{equation}}
\newcommand{\ee}{\end{equation}}
\newcommand{\beaa}{\begin{eqnarray*}}
\newcommand{\eeaa}{\end{eqnarray*}}
\newcommand{\bea}{\begin{eqnarray}}
\newcommand{\eea}{\end{eqnarray}}
\newcommand{\para}{\parallel}

\def\dataVec{\mathbfit{V}}

\def\responseSet{\mathbfss{F}}
\def\regSet{\mathbfss{R}}

\def\srVec{\mathbfit{s}}

\def\noiseVec{\mathbfit{n}}

\begin{document}
\date {}
\title[Bayesian inversion for redshifted 21-cm HI signal]
{A Bayesian analysis of redshifted 21-cm HI signal and foregrounds: Simulations for LOFAR}

\author[A. Ghosh, L\'eon~V.~E.~Koopmans, E. Chapman,
V. Jeli\'{c}]{Abhik Ghosh$^{1}$\thanks{E-mail: ghosh@astro.rug.nl},
  L\'eon~V.~E.~Koopmans$^{1}$, E. Chapman$^2$,
  V. Jeli\'{c}$^{1,3,4}$\\$^{1}$ Kapteyn Astronomical Institute,
  University of Groningen, P. O. Box 800, 9700 AV Groningen, The
  Netherlands\\$^2$Department of Physics \& Astronomy, University
  College London, Gower Street, London WC1E 6BT, UK\\$^3$ASTRON, PO
  Box 2, NL-7990 AA Dwingeloo, the Netherlands\\$^4$Ru{\dj}er
  Bo\v{s}kovi\'{c} Institute, Bijeni\v{c}ka cesta 54, 10000 Zagreb,
  Croatia}

\maketitle
\begin{abstract}

  Observations of the EoR with the 21-cm hyperfine emission of neutral
  hydrogen (HI) promise to open an entirely new window onto the
  formation of the first stars, galaxies and accreting black holes. In
  order to characterize the weak 21-cm signal, we need to develop
  imaging techniques which can reconstruct the extended emission very
  precisely. Here, we present an inversion technique for LOFAR
  baselines at NCP, based on a Bayesian formalism with optimal spatial
  regularization, which is used to reconstruct the diffuse foreground
  map directly from the simulated visibility data. We notice the
  spatial regularization de-noises the images to a large extent,
  allowing one to recover the 21-cm power-spectrum over a considerable
  $k_{\perp}-k_{\para}$ space in the range of $0.03\,{\rm
    Mpc^{-1}}<k_{\perp}<0.19\,{\rm Mpc^{-1}}$ and $0.14\,{\rm
    Mpc^{-1}}<k_{\para}<0.35\,{\rm Mpc^{-1}}$ without subtracting the
  noise power-spectrum. We find that, in combination with using the
  GMCA, a non-parametric foreground removal technique, we can mostly
  recover the spherically average power-spectrum within $2\sigma$
  statistical fluctuations for an input Gaussian random rms noise
  level of $60 \, {\rm mK}$ in the maps after 600 hrs of integration
  over a $10 \, {\rm MHz}$ bandwidth.

\end{abstract}


\section{Introduction}

Observations of redshifted 21-cm radiation from the large scale
distribution of neutral hydrogen (HI) is one of the most promising
probes to study the high redshift Universe. The Epoch of Reionization
(EoR) marks an important milestone in the large scale history of the
Universe. The timing, duration and character of subsequent events
which led to reionization of the Universe contain an enormous amount
of information about the first galaxies and stars
\citep{furla,morales,garrelt}. Evidence from quasar absorption spectra
\citep{becker,fan} and the Cosmic Microwave Background Radiation
(CMBR) \citep{spergel,page,Larson11,Planck15} together imply that the
neutral hydrogen was reionized over a redshift range $6 \le z \le 15$,
but the exact details of how the Universe was slowly reionized, what
were the first sources of radiation in the Universe, how did the EoR
influence structure formation
\citep{barkana,loeb,Pritchard12,Robertson13} are presently
unknown. Through observations of redshifted 21-cm line as a function
of redshift and angular position, we aim to directly measure the HI
distribution in the Universe which will eventually shed light on the
large scale structure formation during this epoch. Currently, several
low-frequency instruments such as Low Frequency ARray
(LOFAR)\footnote{http://www.lofar.org/}, Murchison Widefield Array
(MWA)\footnote{http://www.haystack.mit.edu/arrays/MWA}, Precision
Array for Probing the Epoch of Reionization
(PAPER)\footnote{http://astro.berkeley.edu/~dbacker/eor/}, Giant
Metrewave Radio Telescope
(GMRT)\footnote{http://www.gmrt.ncra.tifr.res.in} are carrying out
observations specifically for detecting red-shifted 21-cm HI
signal. There are also ongoing efforts to build up interferometers
  such as MITEoR \citep{Zheng} and HERA\footnote{http://reionization.org/} which use massive baseline
  redundancy to enable automated precision calibration, reduce system
  noise and cut down on computational cost of the correlator.

Unfortunately, the cosmological 21-cm signal is very faint such that
  no current instrument can detect it directly. It is
  perceived that through a statistical analysis of the fluctuations in the
  21-cm radiation it is possible to observe the high redshift 21-cm
  signal, though currently only upper limits have been placed
  \citep{Paciga,Dillon,Ali15}. Regarding the foregrounds which are
  at least four orders of magnitude larger than the 21-cm signal, it is
  expected that individual point sources can be identified in the
  images and subsequently be removed to a flux level
  depending on the sensitivity of the instrument. The contribution
from the synchrotron emission \citep{shaver} from our Galaxy is
another major diffuse foreground component for detecting the
redshifted 21-cm HI signal.  Other foreground sources include
free-free emission from ionizing halos \citep{oh} and unresolved
extra-galactic sources which could be also large enough compared to
the HI signal. The confusion noise generated from the sea of point
sources below the detection limit can also be a major obstacle in
detecting the HI signal \citep{dimatteo}. For a dipole array such
  as PAPER, which has a wide field of view, \citet{Parsons} have
  developed a delay rate filtering technique which is used to remove the smooth foregrounds and a beam sculpting method depending on the fringe rate of the sources \citep{Parsons15} that can be used to
  select a restricted area near the phase centre to minimize the
  sidelobe contribution coming from sea of point sources away from the
  field centre.

Radio surveys at 408 MHz \citep{haslam}, 1.42 GHz \citep{R82,RR88} and
2.326 GHz \citep{JBN98} have measured the diffuse Galactic synchrotron
radiation at larger than $\sim 1^{\circ}$ angular scales. At
relatively higher frequencies, \citet{giardino01} and
\citet{giardino02} have analyzed the fluctuations in the Galactic
synchrotron radiation using the $2.3 \, {\rm GHz}$ Rhodes Survey and
the $2.4 \, {\rm GHz}$ Parkes radio continuum and polarization
survey. The structure of the Galactic synchrotron emission is not well
quantified at the frequencies and angular scales relevant for
detecting the cosmological 21-cm signal. Though, the Global Sky
  Model (GSM) of \citet{decosta}, which is compiled from publicly
  available total power large-area radio surveys at frequencies $10,
  22, 45, 408 \, {\rm MHz}$ and $1.42, 2.326, 23, 33, 41, 61, 94 \,
  {\rm GHz}$ gives us a representative idea of the diffuse
  Galactic radio emission at a common angular
  resolution of  $5^{\circ}$ and with an accuracy of $1-10 \%$. Recently, \cite{Bernardi09} and
\cite{ghosh150} have analyzed $150 \, {\rm MHz}$ WSRT and GMRT
observations respectively where they find that the Galactic
synchrotron emission is the most dominant foreground at angular scale
$>10$ arcminute after point source subtraction at $10 - 20\, {\rm
  mJy}$ level. A precise characterization and a detailed understanding
of the Galactic synchrotron emission is needed to faithfully remove
foregrounds in 21-cm experiments. The study of the Galactic
synchrotron emission is itself interesting.  This will eventually shed
light on the cosmic ray electron distribution, the strength and
structure of the Galactic magnetic field, and the magnetic turbulence
\citep{Waelkens,Lazarian,iacobelli13}. It is also interesting to point
out that Galactic polarized emission can mimic the EoR signal in
frequency direction and a proper calibration of the instrument is
essential so that the leakages of the polarized to the total signal is
minimal \citep{polarjelic10,geil10,asad15}.

However, interferometers inherently give us a dirty map, which is the
convolution of the true map and direction dependent point spread
functions (PSFs) or synthesized beams corresponding to the sampling
function of the telescope. In general the PSFs are direction-dependent
and typically not invertible \citep{dillon}. Therefore, we need to
develop techniques to accurately characterize (and subsequently
remove) the diffuse foregrounds for every image pixel in the map. The
ultimate aim is to develop calibration and imaging software which will
take into account all the direction dependent effects such as due to
ionosphere \citep{harish14} or the primary beam and direction
independent (such as instrument gains) calibration errors
\citep{sarod}.  We note that in the current generation of imaging
software for LOFAR the calibration solution, in the radio
interferometric measurement equation that describes the instrument and
the ionosphere, is only applied to the model components (mainly
compact sources) in specific directions. In general only a single
directionally independent station based solution is applied to the
residual image. Hence, all images produced by the available imaging
software will include some residual calibration errors which vary from
pixel to pixel and can pose a potential obstacle for EoR detection and
its characterization. Also, the polarized foregrounds leaked from
Stokes Q, U to I maps can mimic the EoR signal if not properly
calibrated. Though, recent results from \citet{asad15} show that in
case of LOFAR a modest polarimetric calibration is sufficient to
ensure that the polarization leakage remains below the EoR signal
inside the Full Width at Half Maximum (FWHM) of the Primary Beam (PB)
of LOFAR. The polarization leakage will also influence the diffuse
foreground characterization and modeling for stokes I images. In this
paper, we examine a Maximum Likelihood (ML) inversion methodology
based on a Bayesian formalism to infer the deconvolved map directly
from the simulated visibilities for the Low Frequency array
(LOFAR). We note that the technique is quite general and can be easily
exported for other interferometers to effectively deconvolve the
effects of instrumental point spread function.

An important point to note here is that the fundamental concept of the
proposed method is based on a Bayesian inference network where we can
incorporate two levels of inference regarding the estimation of the
foreground model. Initially, we fit a foreground model to the data to
determine the free model parameters and in the second level of
inference we can also rank these models. This analysis technique
automatically incorporate `Occam's razor' which ensures that overly
complex foreground models will not be preferred over simpler models
unless the data support them \citep{MacKay}. Regarding the choice of
prior in the Bayesian framework, the proposed technique does not
strongly depend on the inclusion of right priors which will influence
the final outcome, rather many different priors can be tried (such as
identity, gradient or curvature regularization functions) and finally
the data will inform us which is the most appropriate prior based on
the calculated evidence values. We can also maximize the evidence in
light of the data for finding the regularization constant which will
set the strength of the prior in the penalty function.

Based on our simulated ML solutions, we also study how good we are in
recovering the input EoR power spectrum if we only have a diffuse
foreground model with unresolved extra-galactic sources, EoR signal
and Gaussian random noise. We use a non-parametric foreground removal
technique (Generalized Morphological Component Analysis, GMCA) to
remove the foregrounds. GMCA uses a sparsity-based blind source
separation (BSS) technique \citep{bobin} to describe the different
foreground components which are finally combined in different ratios
according to the frequency of the maps to model the foregrounds
\citep{Emma13}.

The paper is organized as follows. In Section 2, we describe our
approach in mathematical formalism. Section 3 discusses the templates
used for generating the visibility data corresponding to the EoR
signal and diffuse foreground emission. In Section 4, we present our
inversion results, while in Section 5 and 6 we describe our foreground
removal results and power spectrum comparison. Finally, in Section 7
we present a summary and possible future extension of the current
work.

\section{Bayesian Inference : Formalism}
\label{formalism}
The fundamental concept of our proposed method is based on a Bayesian
framework \citep{MacKay} which automatically ensures that overly
complex models will not be preferred over simpler models unless the
data support them. The Bayesian framework adopted in this work
automatically incorporates `Occam's razor'. In this section, we give a
general overview of the mathematical formalism which is used to
develop the ML technique. Let us consider a linear system,

\be
\label{eq:lineareq}
$ \dataVec = \responseSet \srVec + \noiseVec$
\ee
 where $\dataVec$ is a vector of data points, $\responseSet$ represents
the response function, $\srVec$ are the source flux parameters that we
want to infer given the data and $\noiseVec$ are the noise in the data
characterized by the co-variance matrix $\mathbfss{C}_{\mathrm{D}}$
(see e.g. \citet{Suyu}). In our case, $\dataVec$ are the measured
visibilities $\mathcal{V}_{\nu}(u,v,w)$, $\responseSet$ is the Fourier
transform kernel $e^{-2\pi i (ul+vm+w(\sqrt{1-l^2-m^2}-1))}$ and
$\srVec$ are the intensities $\frac{I_{\nu}(l,m)}{\sqrt{1-l^2-m^2}}
\times d\Omega_p$ in the sky that we want to infer ($d\Omega_p$ is the
pixel area in solid angle, for details we refer the reader to section
\ref{kernel:ft}).  Assuming the noise to be Gaussian, the probability
of the data given the model parameters $\mathbfit{s}$ is

\be
\label{eq:likelihood}
P(\dataVec | \mathbfit{s}) = \frac {\exp(-E_{\mathrm{D}}(\dataVec | \mathbfit{s}))}{Z_{\mathrm{D}}},
\ee
where 
\bea
\label{eq:ED}
E_{\mathrm{D}}(\dataVec | \mathbfit{s}) &=& \frac{1}{2} \left(\responseSet \srVec - \dataVec \right)^{\mathrm{H}} \mathbfss{C}_{\mathrm{D}}^{-1} \left(\responseSet \srVec - \dataVec \right)
\nonumber \\ & \equiv &
\frac{1}{2}\chi^2
\eea
and $Z_{\mathrm{D}} = (2\pi)^{N_{\rm d}/2} (\det \mathbfss{C}_{\mathrm{D}})^{1/2}$
is the normalization factor.

By definition, the most likely solution ($\mathbfit{s}_{\mathrm{ML}}$)
maximizes the likelihood or minimizes $E_{\mathrm{D}}$. By setting,
$\mathbf{\nabla}
E_{\mathrm{D}}(\mathbfit{s}_{\mathrm{ML}})=\mathbf{0}$, where
($\mathbf{\nabla} \equiv \frac{\mathrm{\partial}}{\mathrm{\partial}
  \mathbfit{s}}$) we find the maximum likelihood (ML) solution being,

\be
\label{eq:MLeq}
\mathbfit{s}_{\mathrm{ML}} = (\responseSet^{\mathrm{H}}
\mathbfss{C}_{\mathrm{D}}^{-1} \responseSet)^{-1}
\responseSet^{\mathrm{H}} \mathbfss{C}_{\mathrm{D}}^{-1}\dataVec
\ee

 In many cases, the problem of finding the most likely
solution that minimizes the half of the $\chi^2$ (here,
$E_{\mathrm{D}}$) is ill-posed and we need to introduce priors to
regularize the solution of $\mathbfit{s}$. 

The posterior probability of the source flux
pixels $\mathbfit{s}$, given the visibility data $\dataVec$, a fixed
form of regularization function $\regSet$ and a level $\lambda$ of
regularization follows from Bayes' theorem,

\be 
\label{eq:posterior}
P(\mathbfit{s}|\dataVec,\lambda,\regSet) = \frac{P(\dataVec | \mathbfit{s}) P(\mathbfit{s}|\regSet, \lambda)}{P(\dataVec|\lambda, \regSet)},
\ee
where $P(\dataVec|\lambda,\regSet)$ is called the \textit{evidence}
which depends on the model parameters $\{\lambda, \regSet\}$. We
choose the model that maximizes the evidence. In
the first level of inference, the most probable (MP) solution maximizes
the posterior, which can be written as,

\be
\label{eq:posterior2}
P(\mathbfit{s}|\dataVec,\lambda,\regSet) =
\frac{\exp(-M(\mathbfit{s}))}{Z_{\mathrm{M}}(\lambda)}, \ee where
$M(\mathbfit{s}) = E_{\mathrm{D}}(\mathbfit{s}) + \lambda
E_{\mathrm{S}}(\mathbfit{s})$, $\lambda$ is the regularization
constant, $E_{\mathrm{S}}$ is called the regularizing function and
$Z_{\mathrm{M}}(\lambda) = \int \mathrm{d}^{N_{\rm s}} \mathbfit{s}
\exp(-M(\mathbfit{s}))$ is the normalizing evidence function. 

In our analysis we choose three most common quadratic forms of
  regularization functions (e.g., zero order, gradient and
  curvature). We subsequently maximize the evidence to fix the optimal
  regularization function and regularization parameter. This allows
  for a fair comparison between any form of regularization. Although
  we don't claim to have found the ``best'' form, we note that for
  reconstructing smooth diffuse foregrounds or the 21-cm signal,
  higher order regularization functions (i.e. gradient or curvature)
  in general work better compared than the zeroth order regularization
  (\citet{Suyu}; see also \citet{Koop05} where these function were also
  used for lensing purposes). Any other form that can be conceived can
  in that case objectively be compared to the ones that we chose to
  test, which are the simplest forms possible. In index and summation
  notation the zero order regularization can be expressed as

\be
\label{eq:ESquad}
E_{\mathrm{S}}(\mathbfit{s}) = \frac{1}{2} \sum_{i=1}^{N_{\rm s}}  s_i^2,
\ee
which tries to minimize the flux at every source pixel.

For gradient and curvature forms of regularization let $s_{i_1,i_2}$
be the source flux at pixel $(i_1,i_2)$, where $i_1$ and $i_2$ range
from $i_1=1,\ldots,N_{\rm 1s}$ and $i_2=1,\ldots,N_{\rm 2s}$, in the two dimensions, respectively. $N_{\rm s}=N_{\rm 1s} N_{\rm 2s}$ is
the total number of flux pixels.  We use the following form of gradient
regularization,

\bea
\label{eq:ESgrad}
E_{\mathrm{S}}(\mathbfit{s}) &=& \phantom{+} \frac{1}{2} \sum_{i_1=1}^{N_{\rm 1s}-1} \sum_{i_2=1}^{N_{\rm 2s}} \left[s_{i_1,i_2}-s_{i_1+1,i_2}\right]^2 
\nonumber \\ & &
+ \frac{1}{2} \sum_{i_1=1}^{N_{\rm 1s}} \sum_{i_2=1}^{N_{\rm 2s}-1} \left[s_{i_1,i_2}-s_{i_1,i_2+1}\right]^2 
\eea
which tries to minimize the difference in the flux values between adjacent pixels on both dimensions.

Similarly, the curvature regularization function that we use is

\bea
\label{eq:EScurv}
E_{\mathrm{S}}(\mathbfit{s}) &=& \phantom{+} \frac{1}{2} \sum_{i_1=1}^{N_{\rm 1s}-2} \sum_{i_2=1}^{N_{\rm 2s}} \left[s_{i_1,i_2}-2s_{i_1+1,i_2}+s_{i_1+2,i_2}\right]^2  
 \nonumber \\ & & 
+ \frac{1}{2} \sum_{i_1=1}^{N_{\rm 1s}} \sum_{i_2=1}^{N_{\rm 2s}-2} \left[s_{i_1,i_2}-2s_{i_1,i_2+1}+s_{i_1,i_2+2}\right]^2
\eea
which minimizes the sum of the curvature in both dimensions.

The most probable solution can be derived from the most likely
solution. For mathematical convenience, we introduce some parameters
$\mathbfss{B}$ and $\mathbfss{C}$ which are the Hessian of
$E_{\mathrm{D}}$ and $E_{\mathrm{S}}$ respectively
($\mathbfss{B}=\nabla \nabla E_{\mathrm{D}}(\mathbfit{s})$ and
$\mathbfss{C}=\nabla \nabla E_{\mathrm{S}}(\mathbfit{s})$). We define
the Hessian of M as $\mathbfss{A}=\nabla \nabla M(\mathbfit{s})$,
which is equivalent to $\mathbfss{A}=
\mathbfss{B}+\lambda\mathbfss{C}$. Following \cite{Suyu},
$\mathbf{\nabla}
E_{\mathrm{D}}(\mathbfit{s}_{\mathrm{ML}})=\mathbf{0}$ gives
$\mathbfit{s}_{\mathrm{ML}} = \mathbfss{B}^{-1} \responseSet^{\mathrm{H}}
\mathbfss{C}_{\mathrm{D}}^{-1} \dataVec$ and the most probable (MP)
solution can be estimated by
$\mathbf{\nabla}M(\mathbfit{s}_{\mathrm{MP}})=\mathbf{0}$, which can
be written as $\mathbfit{s}_{\mathrm{MP}} =
\mathbfss{A}^{-1}\mathbfss{B}\mathbfit{s}_{\mathrm{ML}}$. We note that
the MP solution depends on the regularization constant $\lambda$ since
the Hessian $\mathbfss{A}$ depends on $\lambda$. 

To find the value of the optimal regularization parameter we maximize, 

\be
\label{eq:lamProb}
P(\lambda | \dataVec, \regSet) = \frac{P(\dataVec|\lambda, \regSet) P(\lambda)}{P(\dataVec|\regSet)},
\ee 

Assuming a flat prior in $\log \lambda$, we maximize the evidence in
order to optimize $\lambda$.

For any quadratic form of the regularization function
$E_{\mathrm{S}}(\mathbfit{s})$ we can write the evidence as,

\be
\label{eq:Evid}
P(\dataVec|\lambda, \regSet) = \frac{Z_{\mathrm{M}}(\lambda)}{Z_{\mathrm{D}} Z_{\mathrm{S}}(\lambda)},
\ee
where, 

\be
\label{eq:ZS}
Z_{\mathrm{S}}(\lambda)=\left( \frac{2 \pi}{\lambda} \right)^{N_{\rm s}/2} (\det \mathbfss{C})^{-1/2} e^{-\lambda E_{\mathrm{S}}(\mathbf{0})} , 
\ee
\be
\label{eq:ZM}
Z_{\mathrm{M}}(\lambda)=(2\pi)^{N_{\rm s}/2} (\det \mathbfss{A})^{-1/2} e^{-M(\mathbfit{s}_{\mathrm{MP}})},
\ee

We optimize the log of evidence to find the optimal value of regularization
constant $\lambda$. Using
eqn. \ref{eq:ZS} and \ref{eq:ZM} the log of evidence can be expressed
as,

\bea
\label{eq:EvidFull}
\log P(\dataVec|\lambda, \regSet) &=& -\lambda E_{\mathrm{S}}(\mathbfit{s}_{\mathrm{MP}}) - E_{\mathrm{D}}(\mathbfit{s}_{\mathrm{MP}})  
\nonumber \\ & & - \frac{1}{2}\log(\det\mathbfss{A}) +
\frac{N_{\rm s}}{2}\log\lambda + \lambda E_{\mathrm{S}}(\mathbf{0})
\nonumber \\ & &  + \frac{1}{2}\log(\det\mathbfss{C}) -\frac{N_{\rm d}}{2}\log(2\pi)
\nonumber \\ & & + \frac{1}{2}\log(\det \mathbfss{C}_{\mathrm{D}}^{-1}).
\eea
Here, $N_{\rm d}$ is the number of visibility samples.

\begin{figure*}
\includegraphics[width=120.0mm]{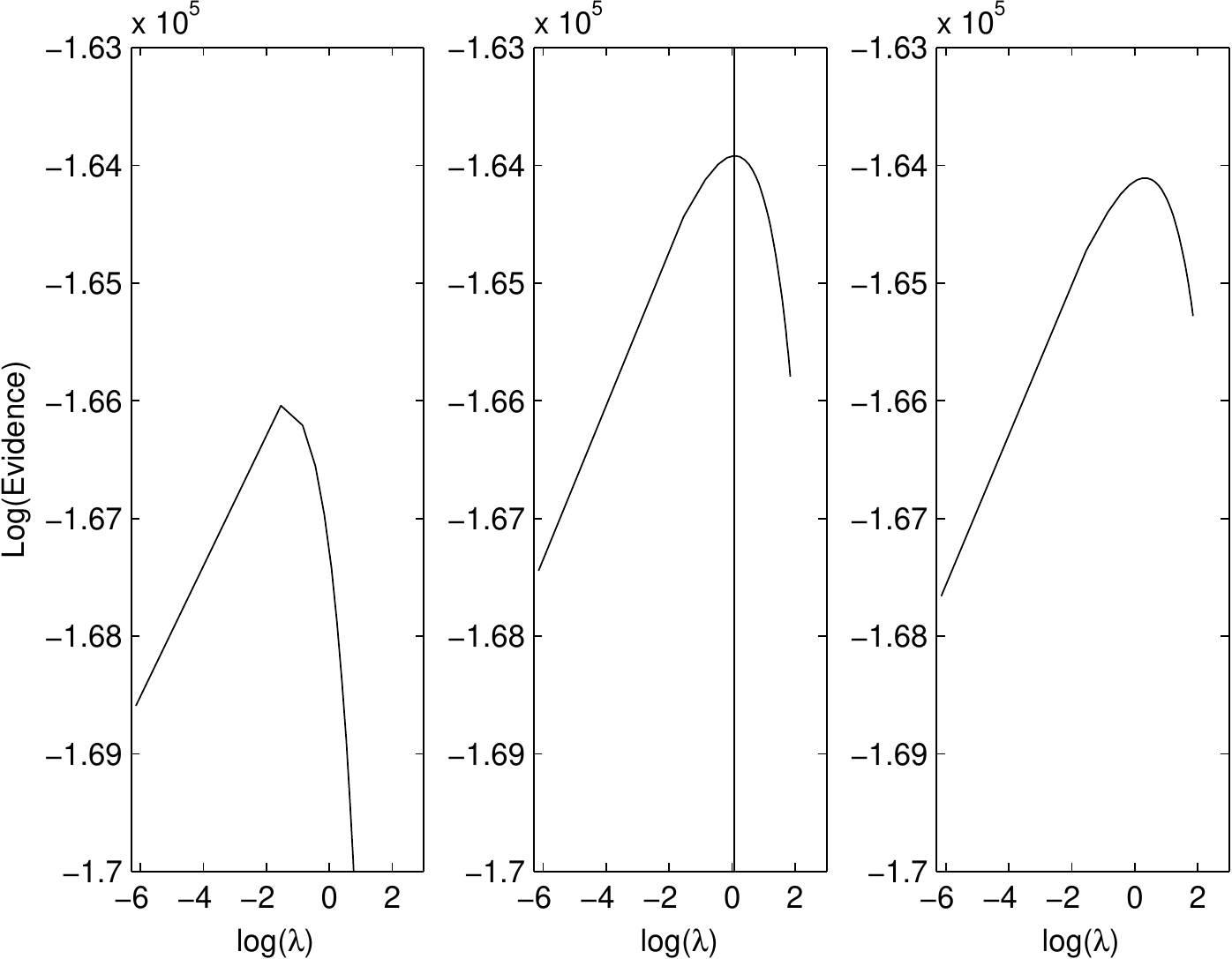}
\caption{From left to right this figure shows the log evidence as a
  function of log of the regularization parameter $\lambda$ for
  identity, gradient and curvature regularization function
  respectively. The log evidence is maximum for gradient
  regularization and the vertical line in the 2nd panel displays the
  peak of the log evidence which is around $\rm{log}(\lambda)=0.0582$, or
  equivalently $\lambda=1.06$.}
\label{fig:LogEvidence}
\end{figure*}

Assuming a flat prior in $\log \lambda$, we maximize the evidence
$P(\dataVec|\lambda, \regSet)$ to find the optimal regularization
constant $\hat{\lambda}$. Solving
$\frac{\mathrm{d}}{\mathrm{d}\log\lambda} \log P(\dataVec|\lambda,
\regSet)=0$ \citep{Suyu}, we get the following non-linear equation for
$\hat{\lambda}$, \be
\label{eq:OptLam}
2 \hat{\lambda} E_{\mathrm{S}}(\mathbfit{s}_{\mathrm{MP}}) = N_{\rm s}
- \hat{\lambda} \mathrm{Tr} (\mathbfss{A}^{-1} \mathbfss{C}), \ee
where, $N_{\rm s}$ are the number of sky flux pixels. We note that
solving this non-linear equation is computationally intensive for
inversions on a large grid and we find that $\hat{\lambda} =
1/\sigma^2$, where $\sigma^2$ is the noise variance in the data, is a
good choice for setting the value of regularization
constant. Following \citet{Suyu}, we expect M evaluated at the optimal
$\lambda$ value to be equal to half of the number of data points
($M(\mathbfit{s}_{\mathrm{MP}}) \sim N_{\rm d}/2$). We find that
$\hat{\lambda} = 1/\sigma^2$ follows this quite well and the
difference is close to the $1\%$ level. We emphasize here that in our analysis we solve eqn. \ref{eq:OptLam} explicitly to find the optimal
  $\lambda$ value with different regularization functions namely,
  identity, gradient or curvature regularization. We also
note that regarding the choice of the functional form of prior in this
framework, our technique does not heavily depend on the inclusion of
very precise priors (indeed a genuine worry in many Bayesian methods,
but one that is mitigated in the framework laid out by \cite{MacKay}
which we use here) which will influence the final outcome. Rather,
different prior families can quantitatively be compared
(e.g. identity, gradient or curvature regularization functions) and
the data inform us which is the most appropriate prior (both shape and
strength) based on the calculated evidence (i.e. the marginalized
posterior). For ranking different models the regularization
  factor $\lambda$ is a nuisance parameter which ends up being
  marginalized. In general the distribution of $\lambda$ is sharply
  peaked and we can approximate $P(\lambda | \dataVec, \regSet)$ by a
  delta function peaked at its most probable value $\hat
  \lambda$. Hence, the evidence can be approximated by
  $P(\dataVec|\hat \lambda, \regSet)$. In our case, using the
  non-gridded visibilities we find that the gradient regularization
  works best with peak values of log evidence at the optimal $\lambda$
  value of $-166043$, $-163919$, $-164109$, respectively, for
  identity, gradient and curvature regularization functions (Figure
  \ref{fig:LogEvidence}). Based on the Bayes factors, there is a very
  strong evidence towards choosing the gradient regularization over
  the identity or the curvature forms of regularization \citep{Jeffreys,Kass}. For the
  rest of our analysis we therefore choose the gradient
  regularization function.

\begin{figure*}
\includegraphics[width=120.0mm]{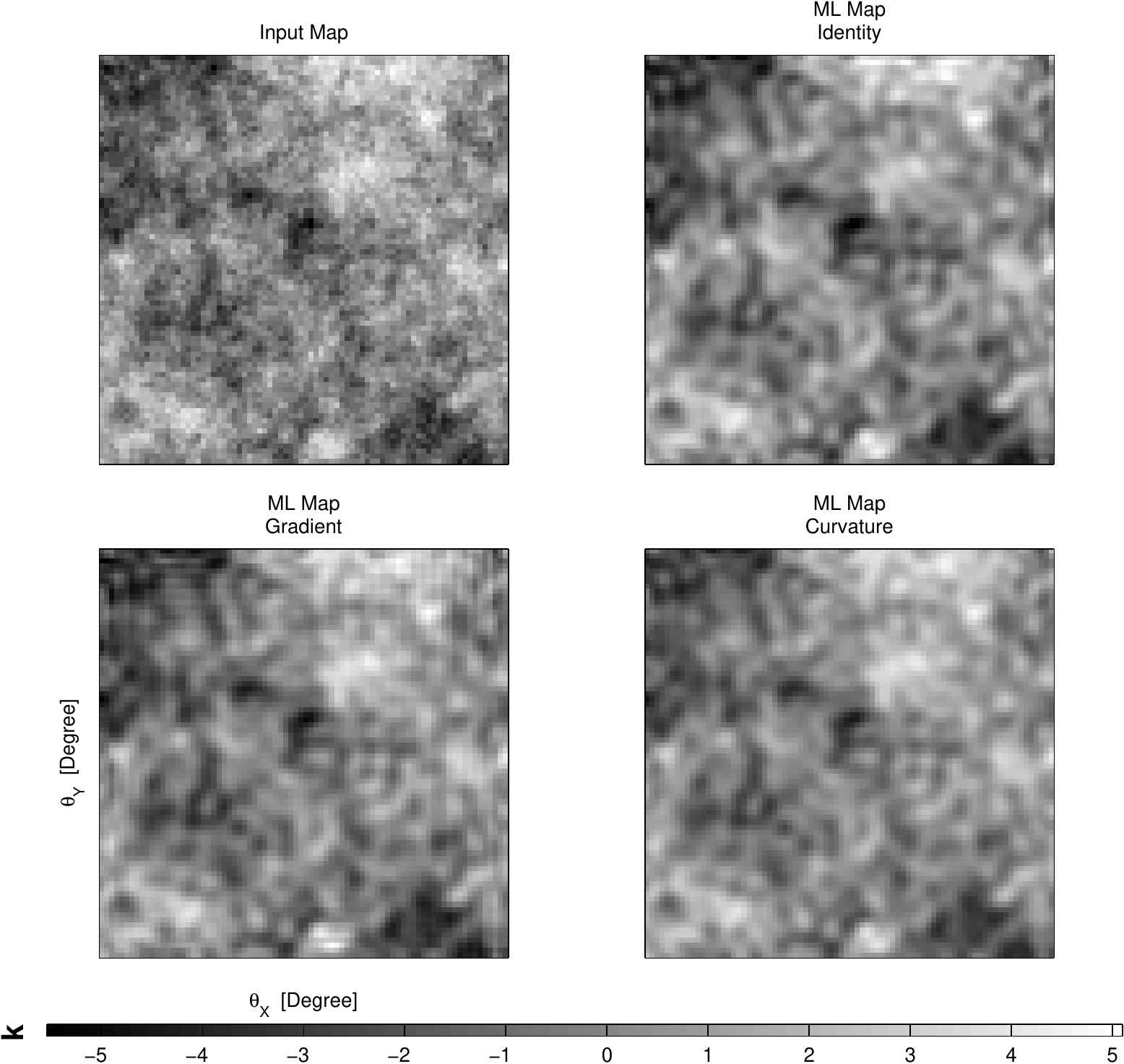}
\caption{The top left panel shows the input map without any noise. The
  remaining panels show the ML inverted maps corresponding to
  identity, gradient and curvature regularization functions
  respectively. Each panel covers a region of $2^{\circ} \times
  2^{\circ}$.}
\label{fig:MPReg}
\end{figure*}

In Figure \ref{fig:MPReg}, we show the ML maps corresponding
to different regularization functions. Even though the ML maps look
similar, based on the evidence values, we decided to use curvature
regularization for rest of our analysis. We note that Occam's razor is
implicit in our evidence optimization. If the model parameter space is
overly-large (for small values of $\lambda$), Occam's razor penalizes
such an overly-flexible model. On the other extreme, for overly-large
values of $\lambda$, the model can no longer fit to the data as the
model parameter space is restricted to a limited region. The optimal
$\lambda$ value lies somewhere in between these two extremes which
ensures that overly complex models will not be preferred over simpler
models unless the data support them. Although, it is in principle
possible to infer $\sigma$ as well, in this paper we assume that we
can infer it independently.

\section{Simulated data templates}

In this section we describe briefly how the data is generated for the
ML analysis. We consider templates for the EoR signal and diffuse
foreground emission from which the visibility samples were generated
and then we added Gaussian random noise to the real and imaginary part
of these visibilities.

\subsection{EoR Signal}

We have used the semi-analytic code 21cmFAST
\citep{Mesinger07,Mesinger11} to simulate the EoR signal. Being a
semi-analytical code 21cmFAST treats physical processes with
approximate methods, but apart from the smaller scales ($< 1 {\rm
  Mpc}$) the results tend to agree well with the hydro-dynamical
simulations of \citet{Mesinger11}. We note that the EoR signal
simulation used here is the same as in \citet{Emma12}. We run the code
using the standard parameters from seven year Wilkinson Microwave
Anisotropy Probe (WMAP) Observations,
($\Omega_{\Lambda},\Omega_m,\Omega_b,n,\sigma_8,h$)=(0.72,0.28,0.046,0.96,0.82,73)\citep{komatsu11}. We
initialized the simulation with $1800^3$ dark matter particles at $z =
300$. The code uses linear perturbation theory to evolve the initial
density and velocity fields to the redshifts of the EoR.  The velocity
field used to perturb the initial conditions were formed on a grid of
$450^3$ and then interpolated to a finer grid of $512^3$.  21cmFAST
utilizes an excursion set formalism to find dark matter halos. For our
simulation, we set $10^9 \rmn{M_{\odot}}$ to be the threshold for
halos contributing ionizing photons. Once the evolved density,
velocity and ionization fields are obtained, the code computes the
brightness temperature $\delta T_{\rmn{b}}$ box at each redshift based
on the following equation,

\begin{eqnarray}
 \delta T_\mathrm{b} &=& 28\: \mathrm{mK} \times (1+\delta)x_{\mathrm{HI}}\left(1-\frac{T_{\mathrm{CMB}}}{T_{\mathrm{spin}}}\right)\left(\frac{\Omega_b h^2}{0.0223}\right) \, \nonumber \\
 && \sqrt{\left(\frac{1+z}{10}\right)\left(\frac{0.24}{\Omega_m}\right)}
\end{eqnarray}
where the brightness temperature fluctuation $\delta T_\mathrm{b}$ is
detected as a difference from the background CMB temperature
$T_{\mathrm{CMB}}$ \citep{field58,field59,ciardi03}, $h$ is the Hubble
constant in units of $100\: \mathrm{km s}^{-1}\:\mathrm{Mpc}^{-1}$,
$x_{\mathrm{HI}}$ is the neutral hydrogen fraction and $\Omega_b$ and
$\Omega_m$ are the baryon and matter densities in critical density
units. For simplicity, we ignore the gradient of the peculiar velocity
whose contribution to the brightness temperature is relatively small
\citep{Ghara14, Shimabukuro}. We also neglected spin temperature
fluctuations by assuming $T_{\rmn{spin}} \gg T_{\rmn{CMB}}$, i.e. the
neutral gas has been heated well above the CMB temperature during EoR
epoch \citep{Pritchard08}.

\subsection{Diffuse foregrounds}

The diffuse foreground model used in this paper is described in great
detail in \citet{Jelic08,Jelic10}. These simulations include Galactic
Diffuse Synchrotron Emission (GDSE), Galactic localized synchrotron
emission, Galactic diffuse free-free emission and unresolved
extra-galactic foregrounds. The Galactic Diffuse Synchrotron Emission
(GDSE) is produced by cosmic ray electrons in the magnetic field of
the Galaxy \citep{ginzburg}.  The intensity and the spectral index of
the GDSE are modeled as Gaussian random fields. The Gaussian fields
are generated assuming a power law spatial power spectrum with 2D
index of $-2.7$. The GDSE is modeled as a power law as a function of
frequency with a spectral index of $-2.55 \pm 0.1$ \citep{shaver}. The
amplitude of the brightness temperature is set around to $253$ K at
$120$ MHz.

Another part of the synchrotron emission comes from the galactic
localized supernovae remnants (SNRs). Along with the GDSE SNRs make up
70 $\%$ of the total foreground emission after point sources are
removed. Inside the observational window of 10\degr $\times$ 10\degr
$\,$ eight SNRs were randomly placed. We note that due to large
computational requirement for our current ML inversion we have
restricted our FoV to 4\degr $\times$ 4\degr $\,$ which corresponds to
the FWHM of the LOFAR primary beam. The SNRs are modeled to be
extended disks and their angular size, flux density and spectral index
are randomly chosen from the observed SNR catalog of D.A. Green (A
Catalogue of Galactic Supernova Remnants, 2006 April version
\footnote{https://www.mrao.cam.ac.uk/surveys/snrs/}).

The free-free emission is generated through the thermal bremsstrahlung
radiation from our galaxy and external galaxies from the diffused
ionized gas. It is also modeled as a Gaussian random filed similar to
GDSE, but with a different frequency spectral index fixed to -2.15
\citep{tegmark,Santos05}. Compared to other foreground components, the
free free emission only contributes 1$\%$ to the total foreground
emission \citep{shaver}, but orders of magnitude higher than the 21-cm
signal.

Unresolved extra-galactic sources, mainly radio galaxies and clusters,
contribute almost $\sim$27 $\%$ of the total foreground emission after
discrete point sources are removed from the map \citep{Jelic08}. The
simulated radio galaxies have a power law spectrum and are clustered
based on a random walk algorithm. The radio clusters are chosen from the
cluster catalogue of Virgo
Consortium\footnote{http://www.mpa-garching.mpg.de/galform/virgo/hubble/}
and using the corresponding relation between the observed mass
luminosity and X-ray radio luminosity functions.

\begin{figure*}
\includegraphics[width=150mm]{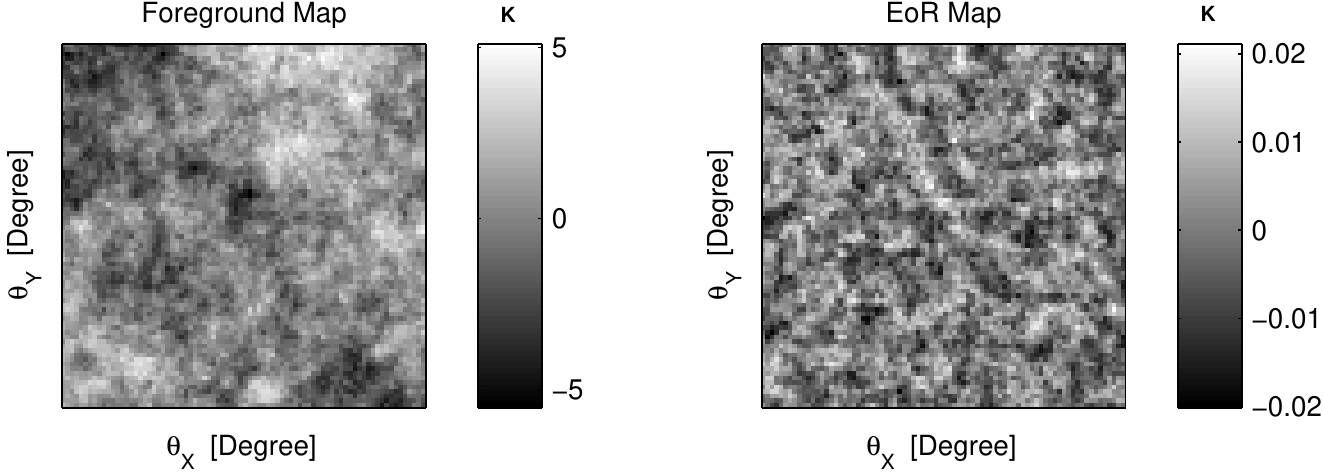}
\caption{This figure shows a slice at $165$ MHz for the diffuse foreground emission and the EoR signal at pixel resolution of 3 arc-min. Note that the mean is subtracted from the maps. Each panel covers a region of $2^{\circ} \times
  2^{\circ}$.}
\label{fig:totfg}
\end{figure*}

Figure \ref{fig:totfg} shows one of the slices at $165$ MHz for the diffuse foreground and the EoR signal that we use in this analysis.

\subsection{Visibility data}
\label{data:vis}

\begin{figure}
\centering
\includegraphics[width=60mm]{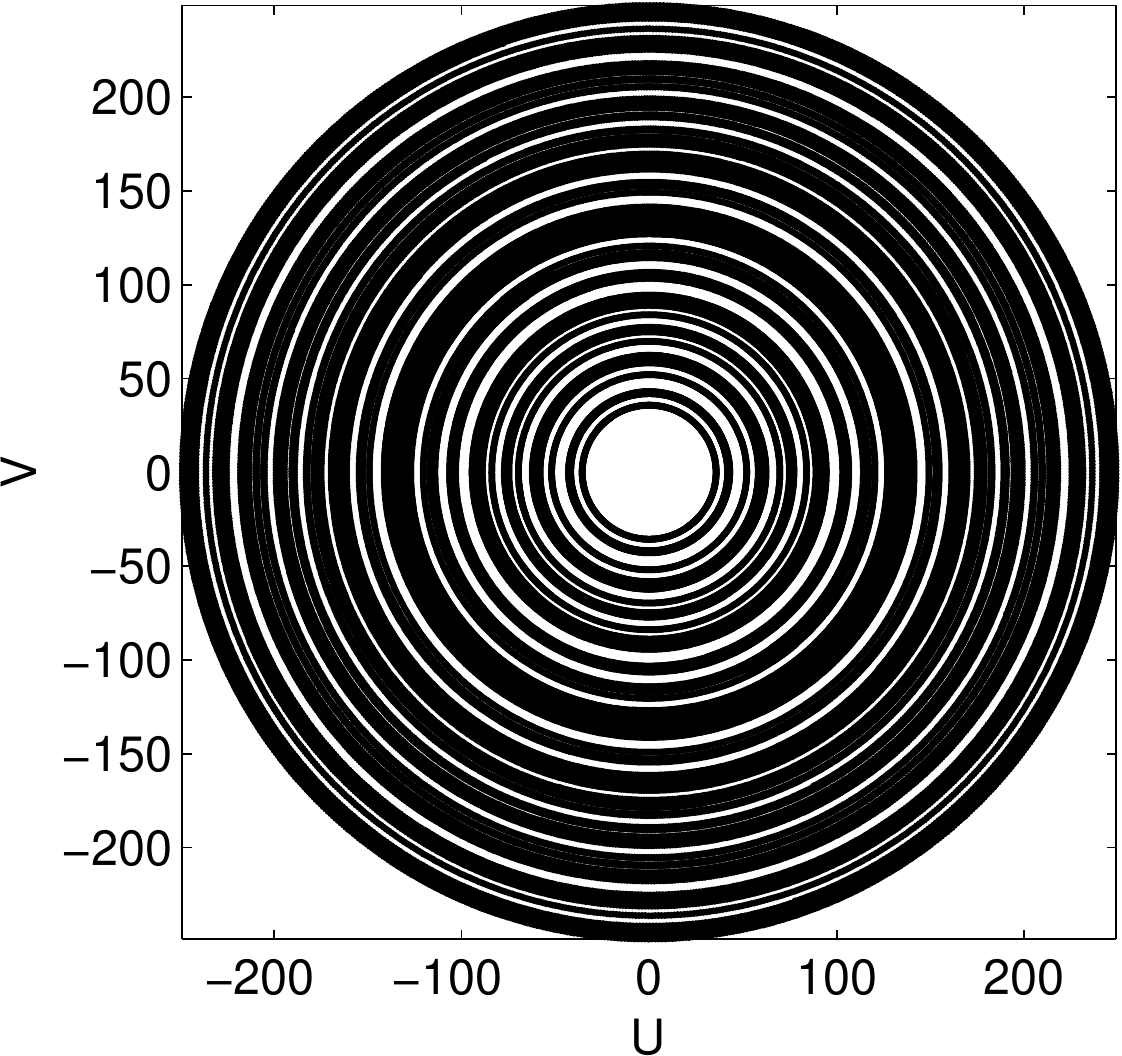}
\caption{The figure shows the $uv$ coverage of our LOFAR
  simulation. We do not show the $w$ component here, but in our
  analysis we take the $w$ component into account. The $uv$ points are
  generated within 30 to 250 $uv$ cut. The $uv$ points and the
  hermitian $uv$ values are shown in this figure. Here $(u,v)$ are the
  antenna separations in wavelength units at a frequency of $165 \,
  {\rm MHz}$.}
\label{fig:uvw}
\end{figure}

We use the LOFAR-HBA antenna positions \citep{Haarlem} to generate the baseline
components $(u, v, w)$ towards the North Celestial Pole (NCP) which
were used to generate the visibility data for our simulation. Figure
\ref{fig:uvw} shows the $uvw$ coverage that we have used for our
simulation. To ensure a manageable data-set, we have sampled
visibilities for each 100 seconds interval for a $uv$ track covering the
full $\pm$ 6 hour angle. The components $u$ and $v$ specify the baseline
coordinates projected on the plane tangent to the NCP, while the third
component $w$ of the coordinate frame points towards the direction of
the NCP. We note that the actual sampling of the $uvw$ points in LOFAR
observation will be higher, but here we restrict our number of
visibility samples as we are limited by the present computational
resources. Note that the number of visibilities is chosen about equal
to the number of independent $uvw$ cells, hence after gridding the
computational effort remains similar even if the number of
visibilities goes up.

The maximum baseline length is restricted to 250 $\lambda$
which corresponds to a maximum wavelength on the sky of $\sim 13$ arcminute. In our inversion, we choose a pixel scale of $\sim 3$ arcminute which ensures that we sample the PSF reasonably well.

\subsection{Fourier Kernel Matrix}
\label{kernel:ft}
A typical radio interferometric array simultaneously measures
visibilities at a large number of baselines and frequency
channels. Ideally, each visibility records a single mode of the
Fourier transform of the specific intensity distribution
$I_{\nu}(\ell,m)$ on the sky. Representing the celestial sphere by a
unit sphere, the component $n$ can be expressed in terms of $(l,m)$ by
$n(l,m)=\sqrt{1-l^2-m^2}$. Then the measured visibility for
monochromatic, unpolarized signal is related with the specific
intensity distribution through the following relation \citep{Perley},

\begin{eqnarray}
\label{eq:Visibility}
\mathcal{V}_{\nu}(u,v,w)&=&\int \frac{I_{\nu}(l,m)}{\sqrt{1-l^2-m^2}} \, \nonumber\\
 && \times e^{-2\pi i (ul+vm+w(\sqrt{1-l^2-m^2}-1))} \mathrm{d}l\mathrm{d}m
\end{eqnarray}

\begin{figure*}
\includegraphics[width=150.0mm]{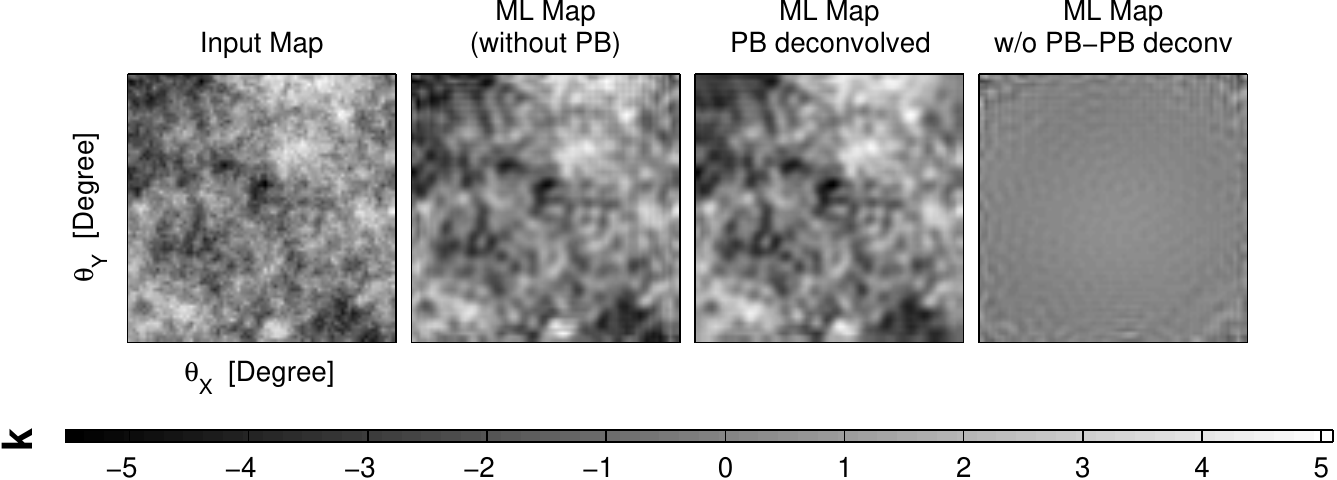}
\caption{From left to right, the $2^{\circ} \times
  2^{\circ}$ panels respectively show the true
  input map without any noise, the inverted map without the any effect
  of the PB, the PB-deconvolved map (where the PB was applied during
  creating the visibility data and the beam was included in the
  Fourier transform kernel to find out the corresponding beam
  deconvolved ML solutions) and the difference between the panel two
  and three. We see that the difference map is very close to zero
  apart from the corner of the field where the PB drops significantly
  from the centre of the field and the noise increases considerably
  which may have some effect on the inversion.}
\label{fig:pbdeconv}
\end{figure*}

We note that without loss of generality (assuming that all stations
have identical beams) we have not incorporated any primary beam
pattern in the visibility definition which in general can be absorbed
into the specific intensity distribution. To test this, we
have introduced a Gaussian Primary Beam (PB) in the sky with a FWHM of 2
  degree within the field of view of 4 by 4 degree and subsequently
  created the input visibility data. During the inversion we absorb
  the PB in the response matrix (as a directionally-dependent
  amplitude to the Fourier kernel) to obtain the corresponding
  ``PB-deconvolved'' true sky. We compare these solutions with the
  solutions presented in this paper where we have not introduced any
  PB in our analysis and we find that these two sets of solutions are
  very similar (Figure \ref{fig:pbdeconv}). We note that here we assume
our inversion grid is in $l, m$ coordinates and the sky phase centre
corresponds to the centre of the grid. Each visibility corresponding
to a $uvw$ value has contribution from all the pixels in the field of
view of $4^{\circ}$ with a pixel resolution of $3$ arcmin. To relate
it with our matrix notation (in Section \ref{formalism}), we recall
that here $\mathcal{V}$ is our data matrix $\dataVec$, the Fourier
kernel, $e^{-2\pi i (ul+vm+w(\sqrt{1-l^2-m^2}-1))}$, is the response
matrix $\responseSet$ and $\frac{I_{\nu}(l,m)}{\sqrt{1-l^2-m^2}}
\times d\Omega_p$ is the model sky parameters that we want to infer
from the visibility data.

\begin{figure*}
\includegraphics[width=150.0mm]{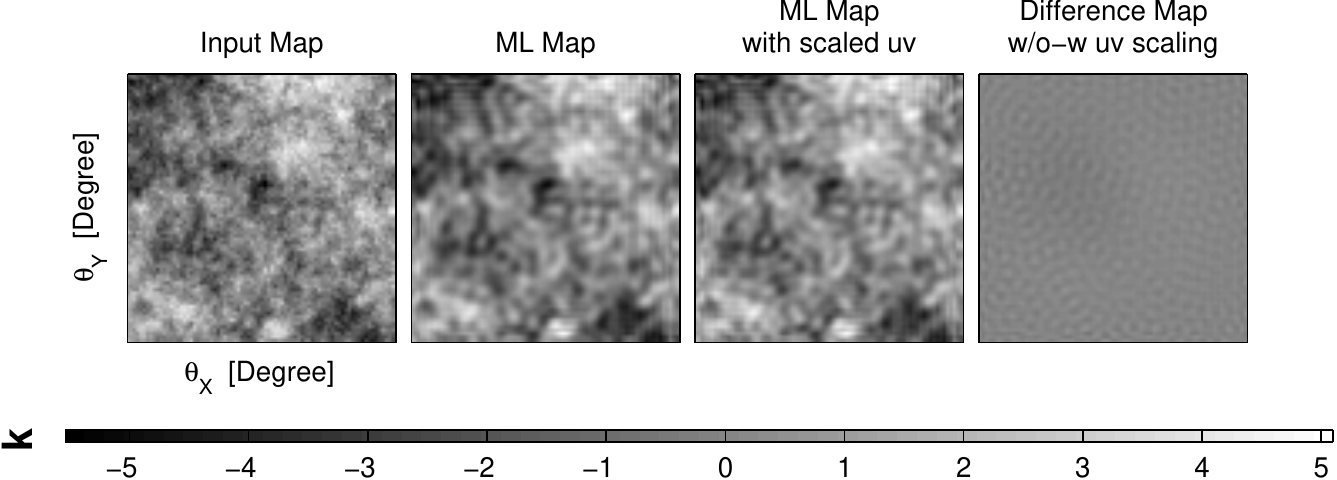}
\caption{The top left panel shows the input map without any noise. The
  rest of the panels respectively show the ML inverted map (w/o PB)
  where the $uv$ values ($U_0$) correspond to the start of the
  frequency band, the 3rd panel show the ML solutions where the $uv$
  values ($U_n$) are scaled according to
  $U_n=U_0(1+n \Delta\nu/\nu_0)$ and then we have selected a
  subsection of visibilities from these scaled $uv$ values which lie
  within a Umax=250 to reconstruct the ML solutions. The last panel
  show the difference between panel 2 and 3 which is mostly close to
  0. Each panel covers a region of $2^{\circ} \times
  2^{\circ}$.}
\label{fig:uvscale}
\end{figure*}

In our simulation, we generate visibilities at exactly the same $(u,
v, w)$ coordinates for all the frequency channels. This ensures that
for a limited sampling of the $uvw$ points the PSF will not vary with
frequency which in principle can be major obstacle for foreground
removal \citep{Bowman09}. We note that in case of a real LOFAR
  observations, where the $uvw$ sampling is quite dense, ensures that
  the PSF variation with frequency can be made very small with proper
  weighting and gridding. LOFAR observations of the NCP show that this
  can be done to a level of better than $30{\rm \,dB}$
  (i.e. $<10^{-3}$). On scales of a few arcminute, this leads to $<1\,
  {\rm mK}$ chromatic effects from any remaining compact source of a
  few mJy. For extended emission that we are simulating this will be
  even less. Therefore, in this simulation paper we choose not to
  scale the uvw values with frequency such that it does not introduce
  any additional effects due to our (currently) limited uvw
  samples. We also did a test where we have generated another sets of
  uvw samples by scaling the uvw values ($U_0$) with frequency as
  $U_n=U_0(1+n \Delta\nu/\nu_0)$ (The number of frequency channels
  $n=20$, channel width $\Delta\nu=0.5 \, {\rm MHz}$, the middle of
  the frequency band $\nu_0=170 \, \rm{MHz}$). Then we only use those
  visibilities within a maximum $uv$ cut of 250 and redo the ML
  analysis. We find the resultant ML solutions are quite similar
  compared to $uvw$ values (those with $U_0$) with no frequency scaling
  (Figure \ref{fig:uvscale}). Moreover, proper filtering of sidelobes
via delay-rate filtering \citep{vedantham12} can mitigate remaining sidelobe
leakage. We therefore believe our current approximation
is justified.

We simulated visibilities over 20 frequency channels of 0.5 MHz
resolution covering a frequency range of 165 to 175 MHz, where the EoR
signal peaks in the simulation \citep{Mesinger07,
  Mesinger11}. The restricted band-width ensures that the evolution of
the 21-cm signal along the line of sight direction due to the light
cone effect \citep{Datta12} will not be a major concern.

In addition to the signal component each visibility also has
a random noise contribution. We added Gaussian noise of rms 60 mK in
the real and imaginary parts of the sampled visibilities to mimic
actual LOFAR observations. This is close to the expected rms level of
noise after 600 hours of LOFAR observation \citep{emma14}. We note
that this noise value is indicative only and it may change somewhat depending
upon the observations.

\section{ML inversion}

In this section we present our ML solutions based on the visibility
data that we generated (Section \ref{data:vis}).  The computational
effort of these linear inversion problem is in general of order $\sim
N_s^3$. For $N_s=80 \times 80$ pixels, this leads to $10^{11-12}$ 
  floating-point operations per frequency channel. To overcome this,
we run this inversion on a parallel 2TB memory machine. Also creating
the matrices from the data itself takes a similar effort.

However, if we have a good estimate of an initial solution (which we
can get in principle via FFTs from the dirty images for different
channels within the frequency bands) then the inversion reduced to a
$\sim N_s^2$, where $N_s$ is the number of sky pixels, where
the `$\sim$' stands for 4-5 iterations in a conjugate gradient (or
quasi-Newtonian) optimization scheme.  Here, because each frequency
slice looks similar to the adjacent one, we use the ML solution of the
previous slice as starting point for the subsequent slice. In this
paper we have done our analysis with both non-gridded and grided data
to assess whether the solutions differ substantially, when creating
the visibility data matrix, the response matrix (the Fourier kernel
which is a filled matrix) and the noise covariance matrix.  For large
number of visibility samples this is quite time consuming and requires
a lot of computer memory. Therefore, we restrict our visibility data
to a size that we expect in real gridded data (about $128^2$ gridded
visibilities) as well as choose to restrict our field of view to 4 by
4 degree with a pixel resolution of 3 arc-minute where (3 arc-minute
and above due to the increase of noise below this scale) we will be
mostly sensitive to detect the 21-cm signal in LOFAR. Overall we think
these choices are fairly reasonable and comparable to grid sampling
and fields of view that are used in real observations.

\subsection{Resulting inversions}

\begin{figure*}
\includegraphics[width=120.0mm]{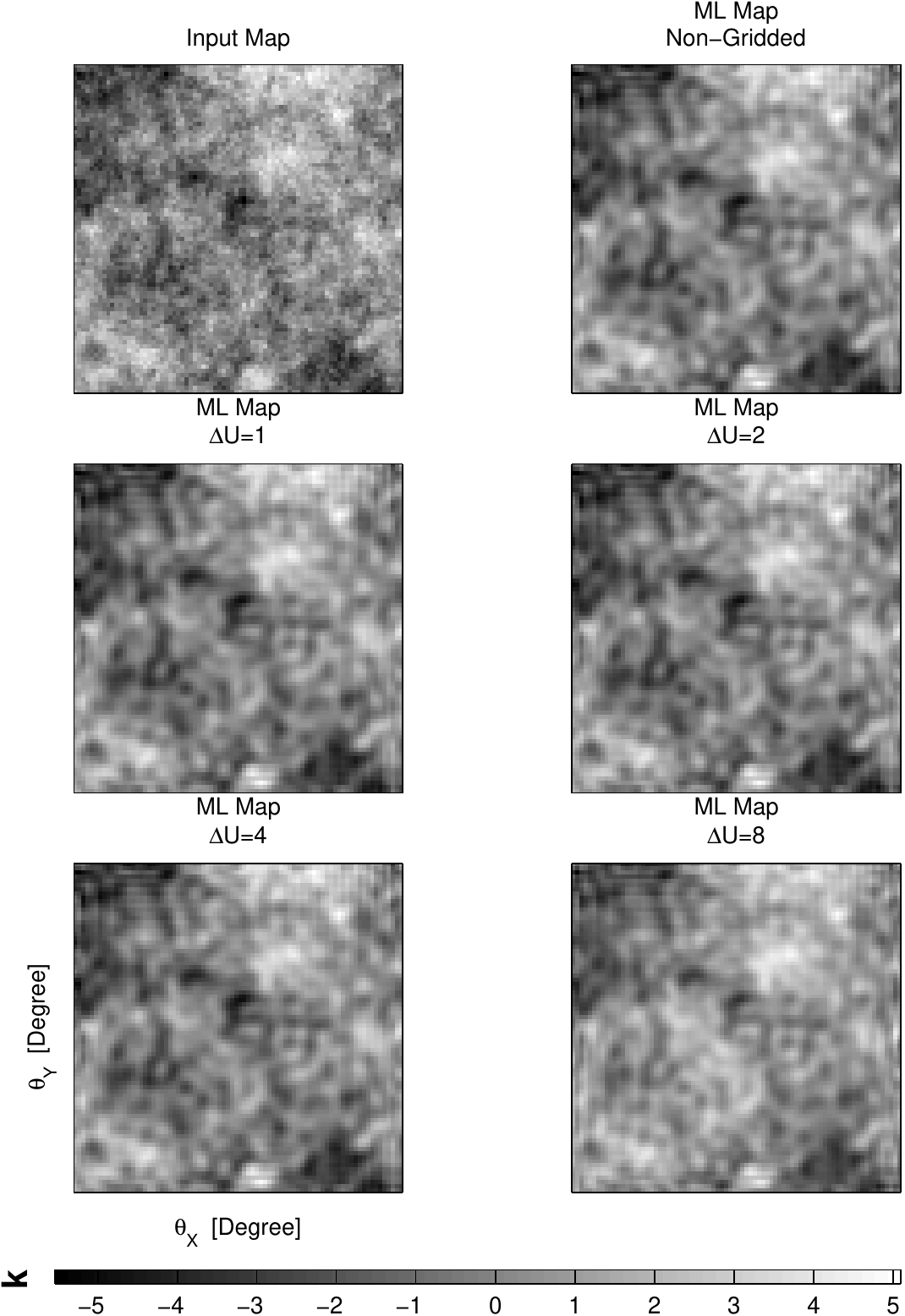}
\caption{The top left panel shows the input map without any noise. The
  remaining panels show the ML inverted maps for non-gridded and
  gridded visibility data with grid size of $\Delta u = $ 1, 2, 4, and
  8 lambda respectively. For the ML inverted maps we have added 60 mK
  of noise to the real and imaginary part of the visibility data, as
  expected for LOFAR after 600 hrs of integration. Each panel covers a
  region of $2^{\circ} \times 2^{\circ}$.}
\label{fig:MP}
\end{figure*}

Figure \ref{fig:MP} shows the ML solutions based on the visibility
data sets that we use. The top most left panel shows the input map,
where the mean is subtracted from the input template maps as
interferometers such as LOFAR is only sensitive to fluctuations in the
sky. Subsequent maps are the results of the ML inversion with
non-gridded and gridded visibility data with grid pixel-size of 1, 2,
4 and 8 lambda respectively. For each of these, we re-optimize the
regularization parameter to maximize the Bayesian evidence. We note
that finer binning is always useful since it provides a better
approximation of the visibility positions in the $uvw$ cube. We averaged
the visibilities according to their $w$ components where the number of $w$
slices is set by $N_w=\lambda B/D^2$, where $\lambda$ is the observing
wavelength, B is the maximum baseline length and D is the station
diameter \citep{synimaging}. The pixel resolution of the ML simulation
is set to 3 arcminute scale sufficient to recover the EoR signal
\citep{Mesinger07}. This also ensures that we sub-sample the PSF
reasonably well (more than 4 times). We note that the ML inversion
maps look smoother relative to the input map because we do not sample
all the small angular scales with our limited $uvw$. Hence, a direct
comparison in the image domain remains difficult because of this
`filtering'.

\subsection{Data and Model Coherency and Foreground Power-Spectrum}

\begin{figure*}
\includegraphics[width=150.0mm]{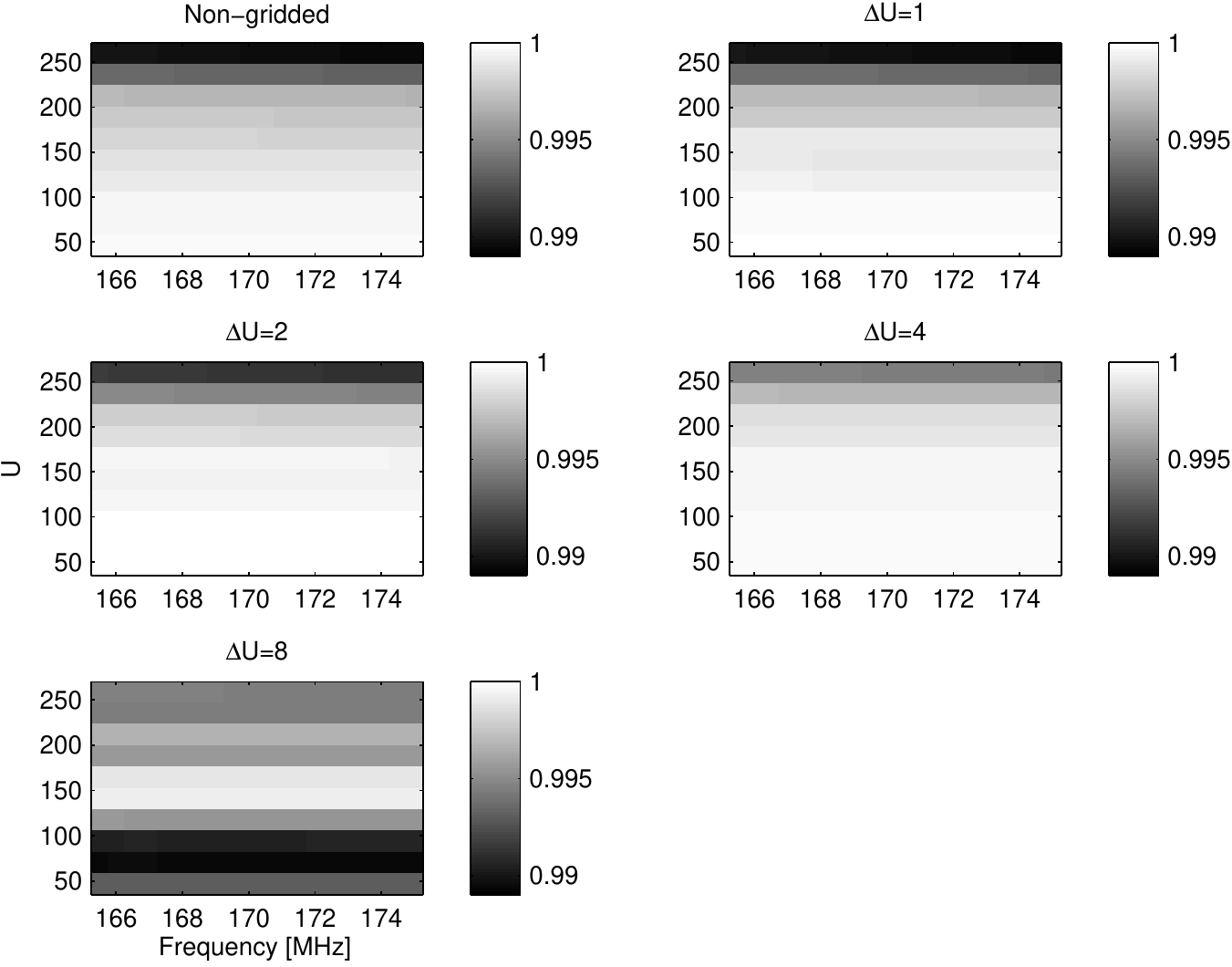}
\caption{The visibility coherency plot between the input data and the corresponding ML non-gridded and gridded visibility data.}
\label{fig:coherency}
\end{figure*}

Next, we computed the visibility coherency matrix between the input
data and reconstructed ML solutions. We
define the coherency matrix as,

\be
\label{eq:cohern}
 V_{\rm {coh}}=\frac{|\langle V_{\rm {data}} V_{\rm{ML}}^* \rangle|^2}{\langle |V_{\rm{data}}|^2 \rangle\langle |V_{\rm{ML}}|^2 \rangle}
\ee
and evaluate this in bins of baselines and across frequency
channels. For a perfect reconstruction, we expect $V_{coh}$ to be 1,
although the effect of noise will lower this value a little (although
the noise is small compare to the foreground flux). The results are
shown in Fig. \ref{fig:coherency} . We notice that apart from $\Delta u=8$, the
coherency matrix is mostly within 0.99-1.0 and this suggests that the
inversion works well when the data is not averaged heavily.

\begin{figure}
\includegraphics[width=75.0mm]{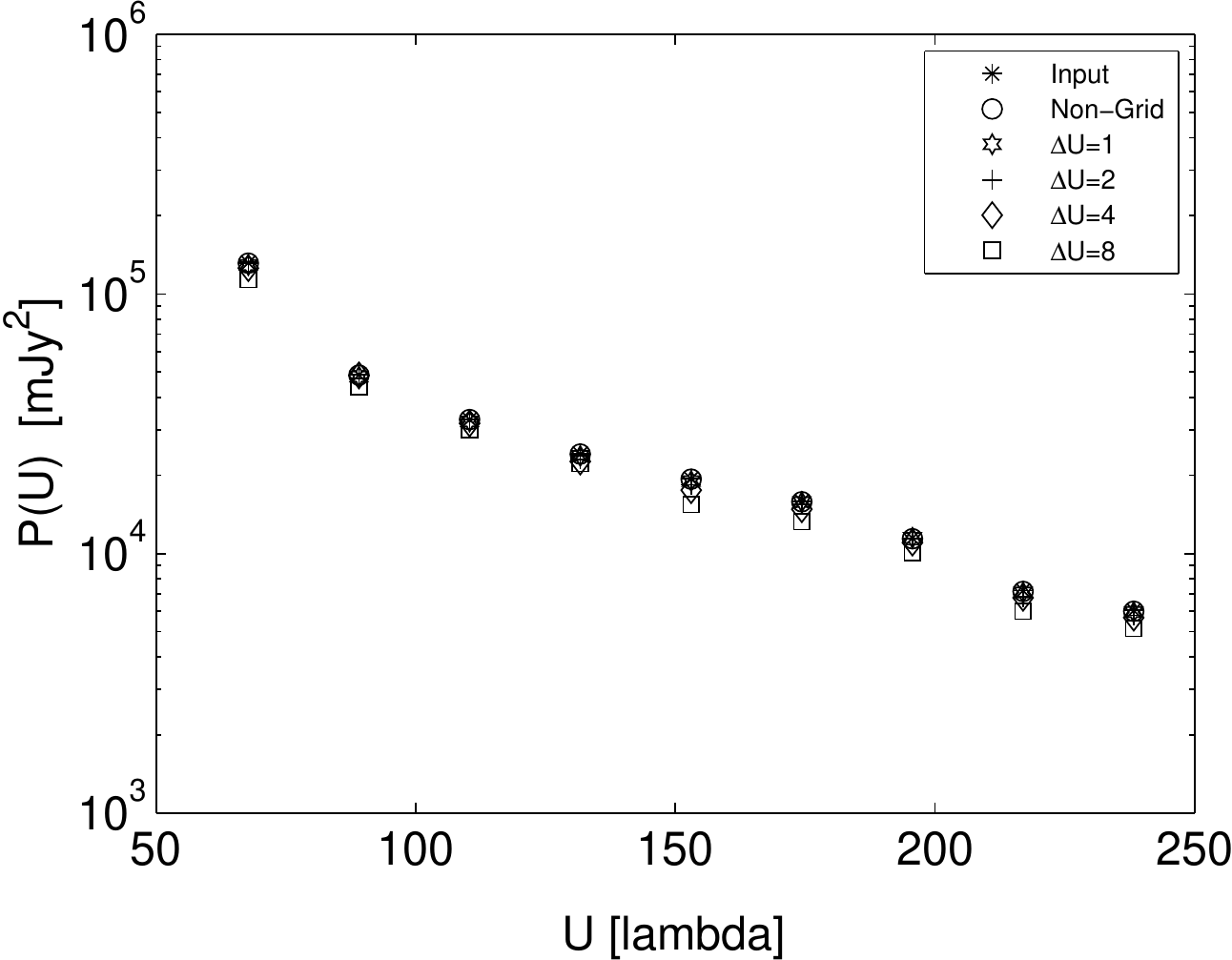}
\caption{The spatial power spectrum corresponding to the input and ML
  maps as function of baseline length. Each panel shows a
  region of $2^{\circ} \times 2^{\circ}$.}
\label{fig:inpmlps}
\end{figure}

We also computed the power spectrum (PS) of the input map and the ML
inverted maps corresponding to the gridded and non-gridded viability data
. We interpolated the PS to common baseline
values so that we can compare different maps. The resulting PS are shown in Figure \ref{fig:inpmlps}. We find that due to the
averaging of the visibilities with different grid sizes there is a
overall drop in the power spectrum on all scales, though the effect is
more pronounced on smaller scales.

In Table \ref{table:redchisq}, we tabulated the reduced $\chi^2$ values
for the non-gridded and gridded visibility data sets. We note that here
we do not compute the evidence values as the visibility data changes each time
for different grid sizes ($\Delta u$). We define the reduced
$\chi^2$ using the number of degrees of freedom as $N_{\rm
  b}-\gamma$, where $N_{\rm b}$ is the number of visibility data
(non-gridded or gridded) that is used in the inversion and $\gamma$ is
the right-hand side of equation (\ref{eq:OptLam}) which determines the
number of ``good" parameters determined by the visibility data.  The
reduced $\chi^2$ commonly determines the goodness of the fit and its
value is close to 1. We note that reduced $\chi^2$ is not strictly 1.0
because Bayesian analysis ensures that evidence is maximized instead
of trying to reach a reduced $\chi^2$ of 1.0. We find that for the
non-gridded visibility data the reduced $\chi^2$ is very close to unity.

\begin{figure*}
\includegraphics[width=150mm]{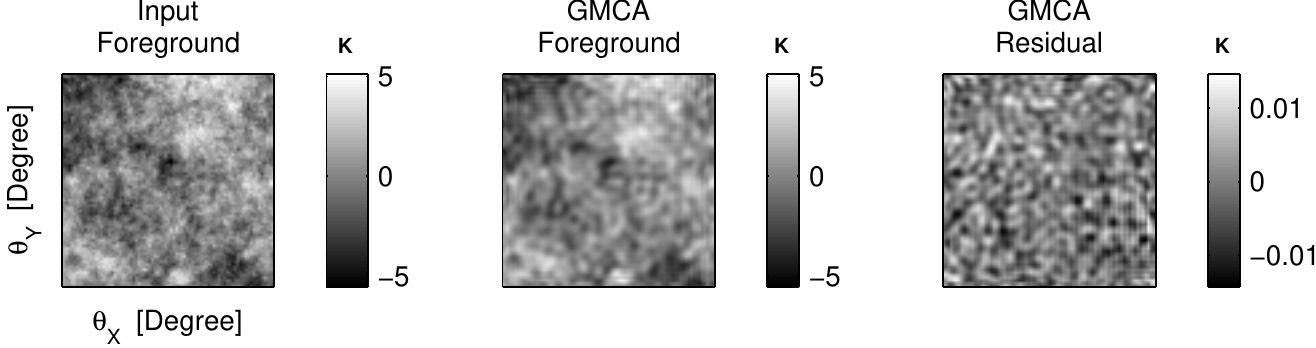}
\caption{A slice at $165$ MHz of the input
  foreground, GMCA reconstructed foreground model and the residual
  after the foregrounds are subtracted. Each panel covers a
  region of $2^{\circ} \times 2^{\circ}$.}
\label{fig:gmcarecons}
\end{figure*}

\begin{table*}
\caption{The reduced $\chi^2$ values evaluated at the regularization constant ($\hat\lambda$) used in the simulation for non-grid and gridded visibility data for different grid sizes ($\Delta u$). The reduced $\chi^2$ is defined as $2E_{\mathrm{D}}/(N_{\rm b}-\gamma)$, where $\gamma = N_{\rm x} - \hat{\lambda} \mathrm{Tr} (\mathbfss{A}^{-1} \mathbfss{C})$ is the right-hand side of eqn. \ref{eq:OptLam}.}

\begin{tabular}{|c|c|c|c|c|c|}
\hline
Gradient & Non-Grid & Grid ($\Delta u=1$) & Grid ($\Delta u=2$) & Grid ($\Delta u=4$) & Grid ($\Delta u=8$) \\
Regularization & & & & & \\
\hline
$\chi^2$ &1.046 &1.07 &1.14 &1.83 &53.21 \\
\hline
\end{tabular}
\label{table:redchisq} 
\end{table*}

Up to $\Delta u=2$, the reduced $\chi^2$ value remains near one after
which it increases. Based on this we decided to carry on our
subsequent analysis with the visibility data which has been gridded
with $\Delta u=2$, as a good compromise between computational speed in
the ML inversion and evidence maximization and goodness of fit. We
note that ML inversion maps contain still the FG and EoR signal and
that the noise in this process is largely filtered. To what level this
has succeeded will be tested in the subsequent section.

\section{Foreground removal}

Low frequency radio observations suggest that the foregrounds are
several orders of magnitude larger than the redshifted 21-cm signal
\citep{Bernardi09,Bernardi10,Ghosh11,ghosh150}. Therefore, removing
the foregrounds is possibly the biggest challenge for detecting the
faint cosmological 21-cm signal. The foregrounds are expected to have
smooth spectra, while the 21-cm signal is line emission which varies
rapidly with frequency (although possibly less rapidly
spatially). This property of the HI signal holds the promise of
allowing us to separate the signal from the foregrounds.

A possible line of approach is to represent the sky signal as an image
cube and for each angular position use polynomial fitting to subtract
out the smooth component of the sky signal that varies slowly with
frequency
\citep{Santos05,Wang06,McQuinn06,Jelic08,Liu09,Liu2nd09,Petrovic11}. The
residual sky signal is expected to contain only the 21-cm signal and
noise. We note that all these parametric foreground fitting techniques
heavily depend on the prior knowledge of the foreground structure
which is largely unknown at the spatial resolution of the low frequencies of
interest. \citet{Liu09} also showed that this method of foreground
removal has problems which could be particularly severe on large
baselines if the $uv$ sampling is sparse. On the other hand,
non-parametric methods do not assume any specific form of the foregrounds
and use the data to decide on the foreground model
\citep{Harker09,Emma12,Emma13}. Here, we use the
Generalized Morphological Component Analysis (GMCA) which is a blind
source separation technique (BSS) which assumes that a
wavelet basis exists in which spectrally smooth foreground components are
sparsely represented with minimal basis coefficients \citep{Emma13} that can be separated from the 21-cm signal.

GMCA utilizes a component separation technique to define the
foregrounds in order to subtract them, treating the 21-cm signal as
noise. We model the foregrounds using four components which refers to
the number of foreground contributions that can be described by unique
sparse descriptions. We note that as more components are added to the
model we might expect the 21-cm signal itself to leak into the foreground
model, although 4 to 5 components seem to work well in recovering the
21-cm signal \citep{Emma13}. In figure \ref{fig:gmcarecons} we show
one of the slice of the reconstructed foreground model using GMCA and
the residual solution. We find an average correlation of $> 70 \%$ with
the true input foreground map and the GMCA reconstructed foreground
model across the $10$ MHz bandwidth.

\section{Denoising and Power-spectra}

\begin{figure}
\includegraphics[width=80.0mm]{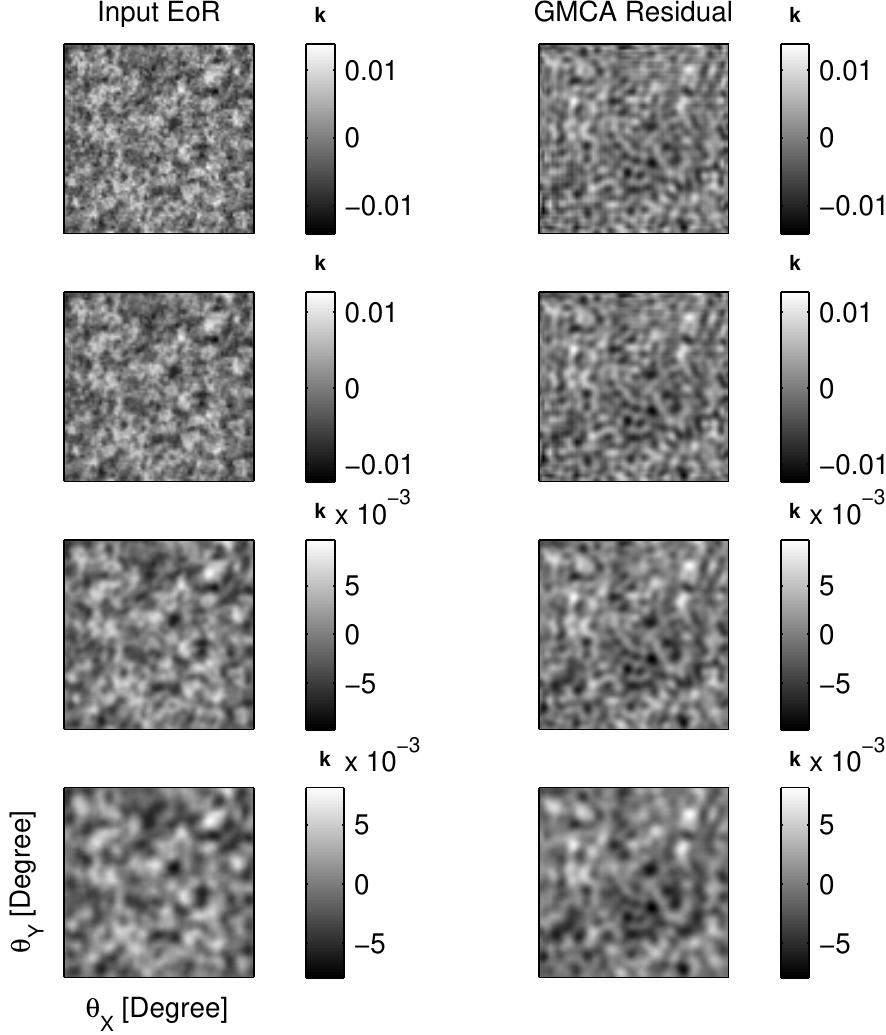}
\caption{The left panel shows the input eor maps and the right panel
  shows the GMCA residual maps. In the top row no smoothing has been
  applied and in the subsequent rows we applied smoothing with sigma =
  0.5, 1 and 2 respectively for a Gaussian filter with size [5 5]
  pixels. Each panel covers a region of $2^{\circ} \times 2^{\circ}$.}
\label{fig:eorgmcares}
\end{figure}

\begin{figure}
\includegraphics[width=80.0mm]{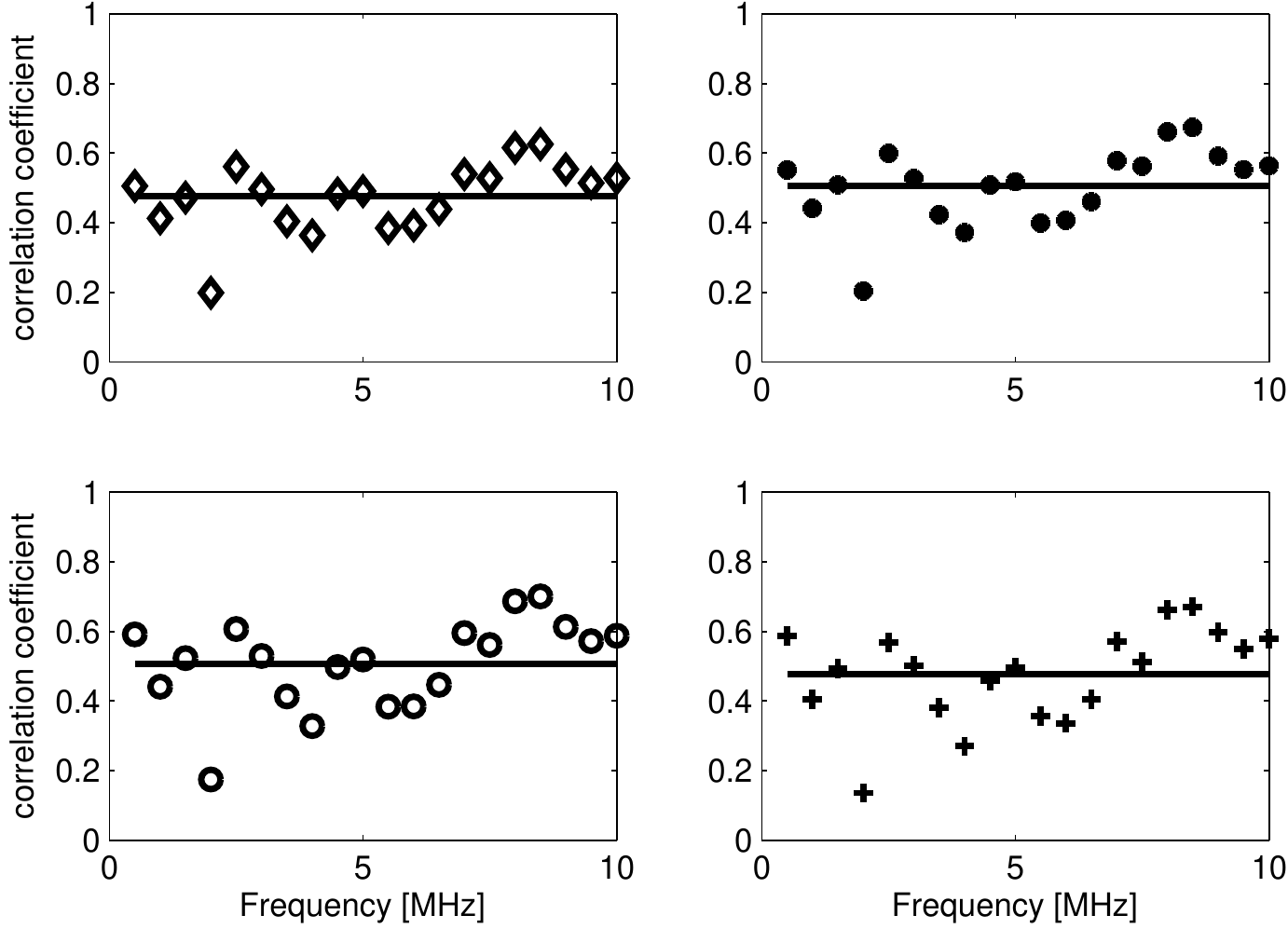}
\caption{The correlation co-efficient between the
  GMCA residual and input EoR as a function of frequency. The black
  horizontal line shows the mean of the correlation. We choose the
  Gaussian filter with size [5 5] pixels and different sigma values.
  Clockwise from the top left panel the figures show the correlation
  coefficient corresponding to no smoothing and smoothing with sigma =
  0.5, 1, 2 respectively.}
\label{fig:corrlall}
\end{figure}

In an ideal situation one would like that after ML inversion the
residual visibilities will be mostly dominated by noise and the ML
solutions will be noise free. We find that we recover most of the
input noises in the residual visibilities and the difference is around
$0.1 \%$ compared to the input noise for the non-gridded case. We note
this is an idealistic situation where we assume that we know the noise
level in the data and we build up the noise covariance matrix using
the same noise values during the ML inversion. The GMCA residual
consists for a large part of the EoR signal and we find a clear
correlation ($\ge 50\%$) between the input EoR signal that we use in
our simulation and the residual GMCA solutions (Figure
\ref{fig:eorgmcares}). The mean correlation between different slices
does not change significantly if we smooth the solution with Gaussian
filter of different widths (Figure \ref{fig:corrlall}), suggesting
that the de-noising seems scale independent. It is interesting to
highlight that if we use a dirty map as an input template to GMCA the
residual map will have contribution from the system noise, the 21-cm
signal and possibly some foreground residuals. Moreover, the success
of the signal recovery requires the noise statistics at different
frequency channels to be determined at an extremely high level of
precision. On the other hand, we find the ML solutions are de-noised
heavily and the residual solutions after GMCA are mostly dominated by
EoR signal. Though, during real observations, the EoR signal detection
can be more complicated where any unmodeled foreground power and any
uncertainties in the bandpass response can prohibit such high level of
precision to be achieved.

\subsection{Power spectrum: Recovered and Input EoR}

This section describes our comparison between the input EoR and the
recovered EoR power spectrum after the ML maps are passed through
GMCA. 

\subsubsection{2D Power spectrum}

\begin{figure}
\includegraphics[width=90.0mm]{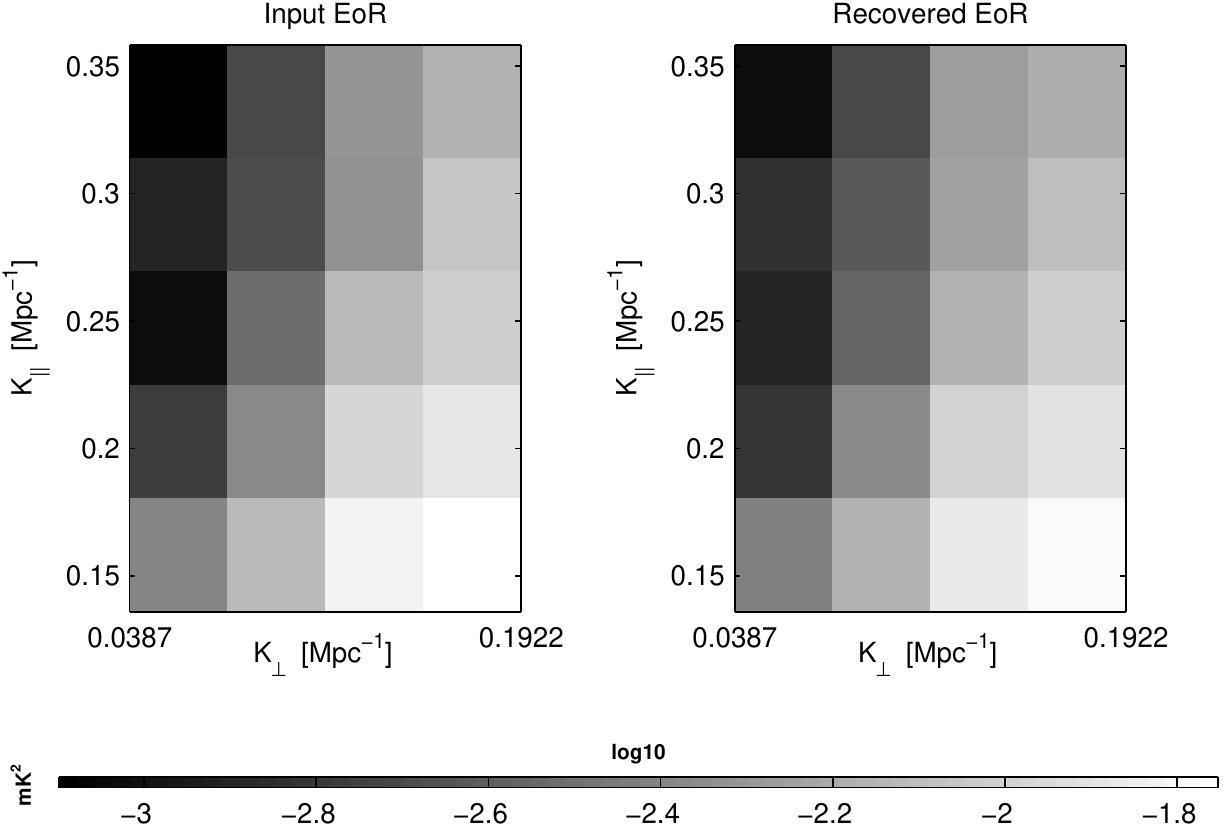}
\caption{This figure shows the cylindrically binned 2D PS for the input and the recovered EoR.}
\label{fig:2dPSVar}
\end{figure}

To test the power spectrum recovery, we compute the cylindrically
binned 2D power spectra from the GMCA residual maps and compare it
with the input EoR power spectrum. We find that the power spectrum is
still dominated by system noise at higher $k_{\perp}$ where the
baseline density drops. In addition, GMCA removes large scale modes
along the frequency direction and this severely suppresses the
power-spectrum on the low $k_{\para}$ scales. Although, this can be
mitigated by running GMCA on wider band-width cubes. In our
comparison, we do not include these high $k_{\perp}$ and low
$k_{\para}$ modes. Figure \ref{fig:2dPSVar} shows the cylindrically
binned 2D power spectra versus the input EoR signal. Note that,
although we did not subtract any noise power spectrum from the
recovered power spectrum, we recover the input EoR power spectrum
quite well. This suggests that the ML solutions are indeed
de-noised. Based on the $k_{\perp}$ scales where we have a reasonable
match with the input EoR power spectrum and using $k_{\perp}=\ell/r$
where $r$ corresponds to the comoving distance at a redshift of 7.35,
we find that we are most sensitive in recovering the EoR signal at an
angular scales of $\sim$ 6 to 17 arcminute.

\begin{figure}
\centering
\includegraphics[width=65.0mm]{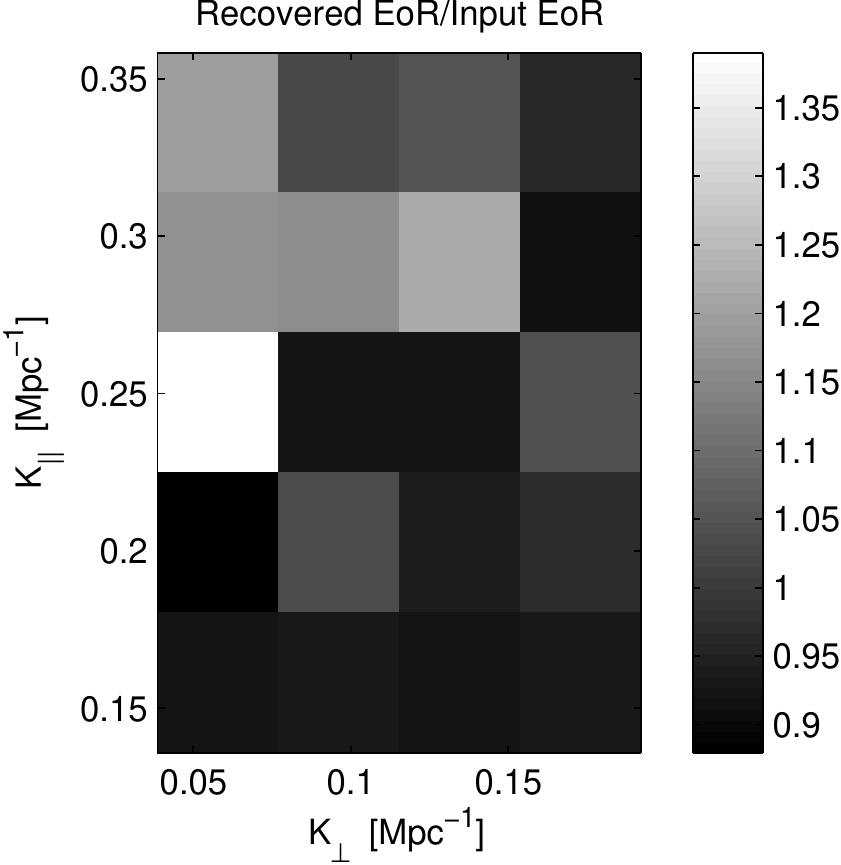}
\caption{This figure shows the ratio of the recovered and input EoR PS.}
\label{fig:psratio}
\end{figure}

To inspect the PS recovery in the $k_{\perp}$ and $k_{\para}$
  plane, we calculated the ratio of the input and the recovered EoR
  PS. Ideally, we expect the ratio of the PS to be close to unity. In
  Figure \ref{fig:psratio} we find that in most of the
  $k_{\perp}-k_{\para}$ plane the ratio plot is close to unity. We
  note in the lower most $k_{\para}$ bin where LOFAR is sensitive
  enough to measure the PS, we find that GMCA removes most of the
  large scale smooth frequency mode and the recovered EoR signal on
  these scales is heavily suppressed. We have therefore avoided these
  low $k_{\para}$ modes in Figure \ref{fig:psratio}. We find in the
  range of k scales shown in Figure \ref{fig:psratio} the PS ratio
  varies from $\sim 0.9 -1.4$, with an overall rms scatter of $0.13$,
  and there are some k modes for which the ratio is slightly
  higher. It is interesting to point out here that the ML inversion
  filters noise spatially and GMCA removes the smooth foregrounds from
  the frequency direction. Therefore, when comparing the recovered and
  input PS we have not subtracted any noise PS from the GMCA
  residuals. This may lead to some extra noise bias in some of the k
  modes where the ratio is larger than 1. In the spherically averaged
  power spectrum (Figure \ref{fig:3DPS}) we find a trend where the
  foregrounds are over subtracted at the small-k modes and there is a
  hint of some extra residual noise on the larger k modes.  Despite
  these differences at the level of several ten percent, we emphasize
  that the results in Figure \ref{fig:psratio} gives us a good
  representation of how well our Bayesian plus GMCA method recovers
  the EoR signal.

\subsubsection{Spherically average power-spectrum}

\begin{figure}
\includegraphics[width=75.0mm]{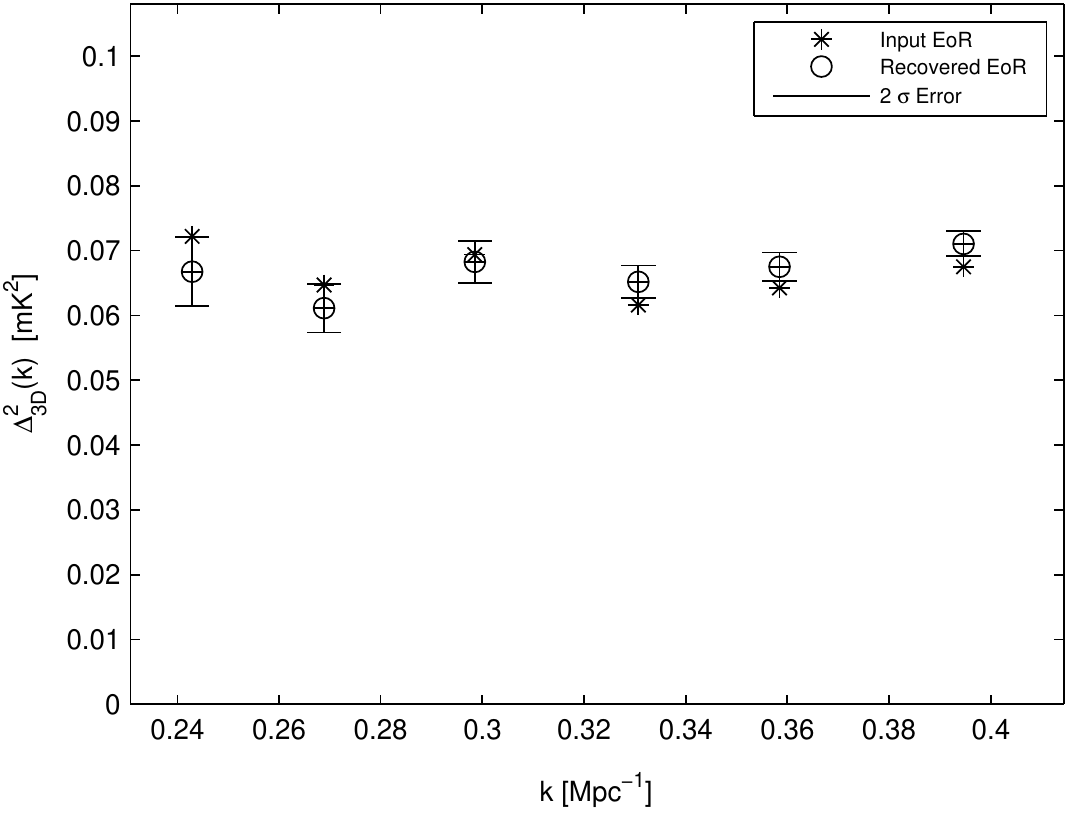}
\caption{The spherically averaged 3D PS for the
  input and recovered EoR signal, along with $2\sigma$ sample variance
  . Note some deviations at the higher k-values, probably due to left-over noise in the image residuals after the ML inversion and GMCA foreground removal. Note that we have not attempted to remove the residual noise power-spectrum.}
\label{fig:3DPS}
\end{figure}

Next, we average the PS in spherical shells and we computed the
spherically averaged dimensionless power spectrum, $\Delta^2({\bf {\rm
    k}})={\rm k^3P}({\bf {\rm k}})/2\pi^2$. In Fig. \ref{fig:3DPS} we
show the input and recovered spherically averaged PS with $2\sigma$
noise fluctuations. The noise contribution only includes sample
variance and the system noise makes a relatively smaller contribution
at these low $k$ modes where we are most sensitive in recovering the EoR
signal. We find the recovered EoR signal is mostly within $2\sigma$
error-bars in the $k$ range shown in Fig. \ref{fig:3DPS}.

\section{Conclusions and Future work}

In this paper, we have introduced a Bayesian framework with spatial
regularization to recover the diffuse foregrounds and the redshifted
21-cm HI signal from a set of calibrated (gridded) visibilities. We
have shown that gradient regularization with optimal regularization
constant which maximizes the evidence works well in reconstructing the
input signal.

Given the large size of the current data sets, it will be a large
computational effort to carry out the inversion directly from the
calibrated visibility data sets, but we have demonstrated that gridded
visibility data can be used effectively in the ML inversion, providing
reduced $\chi^2$ values close to unity. Finer binning gives better
results, since it provides better approximation to the position of the
visibilities in the $uv$ plane. However, binning on scales less than 2
$\lambda$ is not necessary. We note that this is comparable to the
current $(u,v)$ cell-sizes in LOFAR analyzes.

We used the non-parametric foreground removal technique, GMCA
\citep{Emma13}, to remove the smooth diffuse foregrounds. In case of
the cylindrical power spectrum, we can mostly recover the input EoR
signal within a region of $0.14\,{\rm Mpc^{-1}}<k_{\para}<0.35\,{\rm
  Mpc^{-1}}$ and $0.03\,{\rm Mpc^{-1}}<k_{\perp}<0.19\,{\rm
  Mpc^{-1}}$. The high $k_{\perp}$ and low $k_{\para}$ modes beyond
these limits which are accessible in LOFAR observations are mostly
dominated by system noise and the removal of smooth large scale modes
in the frequency direction by GMCA. Scales $0.24\,{\rm
  Mpc^{-1}}<k<0.40\,{\rm Mpc^{-1}}$, are largely (within 2-sigma)
recovered in the spherically averaged 3D input EoR power
spectrum. Although, at higher $k$ scales the residual noise has some
effects and the recovered EoR level is slightly higher compared to the
input power spectrum.

In the near future, we plan to extend our Bayesian imaging technique
for a full Stokes analysis where we will compare the ML solution with
those of the more classical ones and precisely quantify the errors and
correlation between all model parameters. We will also apply our
technique to the real LOFAR data where most of the point sources have
been subtracted and the residual visibilities are dominated by the
unresolved source confusion noises. Another goal is to include
direction dependent (such as ionosphere, beam etc.)  and direction
independent calibration errors (such as gain variation) in predicting
the corresponding sky from the measured data.

\section{Acknowledgment}
AG and LVEK acknowledge the financial support from the European
Research Council under ERC-Starting Grant FIRSTLIGHT - 258942 (PI:
LVEK). VJ would like to thank the Netherlands Foundation for
Scientific Research (NWO) for financial support through VENI grant
639.041.336. AG thanks Saleem Zaroubi, Ajinkya Patil and Khan
M. B. Asad for many useful discussions. We would like to thank the
anonymous referee for providing us with constructive suggestions which
helped to improve the paper.

\footnotesize{
\bibliographystyle{mn2e}

}

\end{document}